\documentclass[11pt]{amsart}

\usepackage{amsmath,amsthm,amssymb,amsfonts}
\usepackage{graphicx,epsfig}

\newtheorem{theorem}{Theorem}[section]

\newtheorem{definition}{Definition}[section]

\newtheorem{proposition}[theorem]{Proposition}
\newtheorem{conjecture}{Conjecture}[section]
\newtheorem{remark}{Remark}

\newcommand{\beq}{\begin{eqnarray}}
\newcommand{\eeq}{\end{eqnarray}}
\newcommand{\beql}[1]{\begin{eqnarray}\label{#1}}
\newcommand{\beqs}{\begin{eqnarray*}}
\newcommand{\eeqs}{\end{eqnarray*}}

\newcommand{\e}{\varepsilon}

\newcommand{\pa}{\partial}

\newcommand{\cc}{{\mathbb C}}
\newcommand{\rr}{{\mathbb R}}
\newcommand{\zz}{{\mathbb Z}}

\newcommand{\dn}{{\rm dn}}
\newcommand{\cn}{{\rm cn}}
\newcommand{\sn}{{\rm sn}}

\begin{document}

\setlength{\baselineskip}{16pt}

\title[The $6$-vertex model]{The $6$-vertex model with fixed boundary conditions.}

\author{Konstantin Palamarchuk}
\address{ Department of mathematics, University of California,
Berkeley, CA 94720, USA}
\email{reshetik@math.berkeley.edu}

\author{Nicolai Reshetikhin}
\address{Department of mathematics, University of California,
Berkeley, CA 94720, USA}
\email{reshetik@math.berkeley.edu}

\begin{abstract}We study the 6-vertex model with fixed boundary
conditions. In the thermodynamical limit there is a formation of
the limit shape. We collect most of the known results about the
analytical properties of the free energy of the model as the
function of electric fields and study the asymptotical behavior
near singularities. We also study the asymptotic of limit shapes
and the structure of correlation functions in the bulk.
\end{abstract}

\maketitle

\begin{center}

\today
\end{center}

\tableofcontents

\section{Introduction}

Ising and dimer models were among the first models in
two-dimensional statistical mechanics for where the partition
function and for some of the correlation functions were computed
explicitly in terms of Pfaffians of certain matrices. For this
reason both of these models can be regarded as theories of
Gaussian discrete two dimensional fermi field. The Ising model was
solved by Onsager and for dimer models the Pfaffian solution was
found by Kasteleyn.

The $6$-vertex model, a particular case of which is the ice model
is interesting for a number of reasons. Physically, it is a model
of ferro- and antiferro- electricity. It has many equivalent
reformulations, one of them (which we will use) describe the
6-vertex configurations on a planar connected simply connected
region in terms of stepped surfaces. One of the combinatorial
reformulations of the 6-vertex model for specific value of
parameters \cite{Kup} is related to alternating sign matrices
\cite{Bres}.

The 6-vertex model generalizes dimer models and can be regraded as
the theory of Gaussian discrete fermions with four fermionic
interaction. The partition function of the 6-vertex model with
periodic boundary conditions was computed in \cite{Lieb} using the
Bethe Ansatz method.

The computation of correlation functions in the $6$-vertex model
is highly non-trivial and to the large degree is still a
challenge. There are two known approaches to this problem based on
the internal symmetry of the model, i.e. on the representation
theory of quantum affine algebras. First approach is based on
determinantal formulae for certain matrix elements  \cite{KBI}.
For some recent results based on this method see \cite{BPZ} and
\cite{CP}. The second approach is based on form-factor formulae
derived in \cite{Sm}. For an overview of this approach see
\cite{JM} and for the latest results see \cite{BJMST}.

The 6-vertex model on a planar simply regions can be reformulated
as the theory of random stepped surfaces. A configuration of
arrows in a 6-vertex model can be interpreted as a configuration
of paths which can be viewed as level corves of a {\it height
function} defining a stepped surface. Gibbs measure on 6-vertex
configurations define a Gibbs measure on stepped surfaces.

For certain class of such Gibbs measures, random surfaces in the
thermodynamical limit develop the limit shape phenomenon
\cite{Shef} also known as the acric circle phenomenon \cite{CEP}.
It means that on macroscopical scale the random surface becomes
deterministic. The fluctuations remain at smaller scale and the
structure of fluctuations may change depending on how singular the
limit shape is at this point. The limit shape phenomenon is
studied in details in dimer models \cite{KO}.

The numerical results from \cite{AR}\cite{SZ} show how limit
shapes develop in a 6-vertex model with domain wall boundary
conditions. In this paper we will focus on the limit shape
phenomenon for the 6-vertex model on a planar connected simply
connected regions with fixed boundary conditions. Computing the
limit shape involves two steps.

Step one is the derivation of the formula for the free energy of
the 6-vertex model as a function of magnetic fields. Unlike the
partition function of dimer models, the free energy can not be
computed explicitly. However, it can be written in terms of the
solution to a linear integral equation. Many analytical properties
of the partition function are known but scattered in the
literature, see for example
\cite{Yang}\cite{Yang1}\cite{SY}\cite{SY1}\cite{LW}
\cite{BS}\cite{Nold}\cite{NK}. We collected most of them in the
section \ref{f-energy} together with some new results.

Step two is to derive and solve the variational principle which
determine the limit shape for given boundary conditions. The free
energy of the model as a function of magnetic fields determine the
functional in the variational problem. Such variational problem
was first introduced for dimer models in \cite{CKP}.  The idea of
using this variational problem in the 6-vertex first appeared in
\cite{Z1} where some interesting partial results were obtained for
correlation functions in the bulk of the limit shape.

The structure of fluctuations near the limit shape is determined
by the asymptotical behavior of correlation functions at smaller
scales. We will discuss this problem for the 6-vertex model in the
last section.

Finally let us mention the special case of domain wall boundary
conditions. These boundary conditions first appeared in the
computation of norms of Bethe vectors \cite{Kor}. The remarkable
fact about them is that the partition function of the 6-vertex
model with these boundary conditions can be written as a
determinant \cite{Ize}. Another remarkable fact is that exactly
these boundary conditions relate the 6-vertex model with
alternating sign matrices \cite{Kup}. The large volume asymptotic
of the partition function of the 6-vertex mode with these boundary
conditions was computed in \cite{KZ}\cite{Z}.

Here is the outline of the paper. In the first two sections we
recall some basic facts about the $6$-vertex model, about its
reformulation it in terms of height functions, and about the
thermodynamical limit in the model. The third section contains the
description of the free energy per site for as the function of
electric fields in thermodynamical limit for periodic boundary
conditions. Some asymptotical behaviors of the free energy are
computed in section 4. This section is a combination of an
overview and original results. In section 5 we study the
asymptotical behavior of the limit shapes near the "freezing
point". In the last section we discuss fluctuations.

We thank C. Evans, R. Kenyon, A. Okounkov, and S. Sheffield for
interesting discussions.  The work of N.R. was supported by the
NSF grant DMS 0307599, by the Niels Bohr initiative at Aarhus
University, by the Humboldt foundation and by the CRDF grant
RUMI1-2622 The work of K.P. was supported by the NSF RTG grant and
DMS 0307599.

\section{The $6$-Vertex Model}

\subsection{The $6$-vertex model}

First, let us fix the notation. A square $N\times M$ grid
$L_{N,M}$ is a graph with $4$- and $1$-valent vertices embedded
into $\rr^2$ such that $4$-valent vertices are located at points
$(n,m), \ n=0,1,\dots, N-1, m=0,1,\dots, M-1$ (see Fig.
\ref{dwbc_conf}) and $1$-valent vertices are located at $(-1,m),
(N,m),m=0,1,\dots, M-1$ and at $(n,-1), (n, M-1), n=0,1,\dots,
N-1$. An edge connecting two $4$-valent vertices is called an {\it
inner} edge and an edge connecting a $4$-valent vertex with a
$1$-valent vertex is called an {\it outer} edge.

\begin{figure}[t]
\begin{center}
\includegraphics{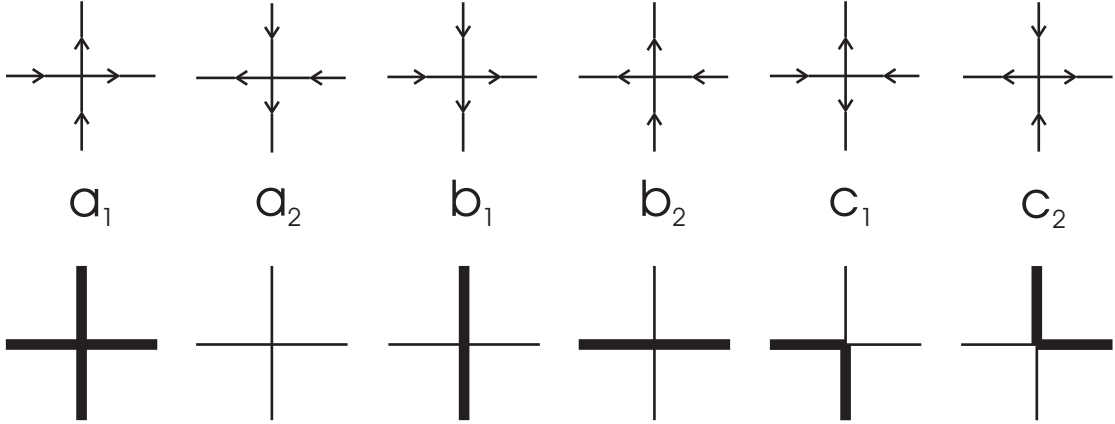}
\end{center}
\caption{The $6$ types of vertices and the corresponding thin and
thick edges configurations} \label{vertices}
\end{figure}

States of the $6$-vertex model on $L_{N,M}$ are configurations of
arrows assigned to each edge (i.e. orientations of $L_{N,M}$).
They satisfy the ice rule: at any vertex the number of incoming
arrows should be equal to the number of outgoing arrows. Six
possible configurations at a vertex are shown on Fig.
\ref{vertices}. Configurations of arrows on boundary edges are
called boundary conditions.

Each configuration of arrows on the lattice can be equivalently
described as the configuration of ``thin'' and ``thick'' edges or
empty and occupied edges shown on Fig. \ref{vertices}. There
should be an even number of thick edges at each vertex as a
consequence of the ice rule. The thick edges form paths. We assume
that paths do not intersect (two paths may  meet at an
$a_1$-vertex). So, equivalently, configurations of the $6$-vertex
model can be regarded as configurations of paths satisfying the
rules from Fig. \ref{vertices}.

To each configuration of arrows on edges adjacent to a vertex we
assign a Boltzmann weight, which we denote by the same letters.
The physical meaning of a Boltzmann weight is
$\exp(-\frac{E}{T})$, where $E$ is the energy of a state and $T$
is the temperature (in the appropriate units). Thus, all numbers
$a_1$, $a_2$, $b_1$, $b_2$, $c_1$, and $c_2$ should be positive.

Choosing the scale such that $T=1$, it is natural to write
Boltzmann weights in the exponential form.
\begin{eqnarray}
&& a_1=e^{-E_1+H+V}, \qquad a_2=e^{-E_1-H-V}, \nonumber
\\
&& b_1=e^{-E_2+H-V}, \qquad b_2=e^{-E_2-H+V}, \nonumber
\\
&& c_1=e^{-E_3}, \qquad c_2=e^{-E_3}, \nonumber
\end{eqnarray}
where $E_1$, $E_2$, and $E_3$ are dimensionless interaction
energies of arrows at different types of vertices, and $H$ and $V$
are dimensionless horizontal and vertical components of the
magnetic field, respectively. In this interpretation arrows are
spins interacting with the magnetic field. We set $c_1=c_2$
because for the types of boundary conditions we will consider the
difference between the number of $c_1$-vertices and $c_2$ vertices
is the same for all states and, therefore, the probability does
not depend on the ratio $c_1/c_2$.

We also use the standard notation
$$
a=e^{-E_1}, \qquad b=e^{-E_2}, \qquad c=e^{-E_3}.
$$
These are the weights of the model when there is no magnetic
field.

The weight of a state is the product of weights of vertices in the
state. The weight of a state on $L_N$ (up to a constant factor)
can be written in terms of energies and magnetic fields as
\[
\exp(-E_1N(a)-E_2N(b)-E_3N(c)+\frac{H}{2}N(hor)+\frac{V}{2}N(ver))
\]
where $N(a)$ is the total number of $a$-vertices, $N(b)$ is the
total number of $b$-vertices, $N(c)$ is the total number of
$c$-vertices, $N(hor)$ is the total number of horizontal edges
occupied by paths, and $N(vert)$ is the total number of vertical
edges occupied by paths.

The partition function is the sum of weights of all states of the
model
$$
Z=\sum_{\rm states}\prod_{\rm vertices} w(\rm vertex),
$$
where $w(\rm vertex)$ is one of the weights from Fig.
\ref{vertices}.

Weights define the probabilistic measure on the set of states of
the $6$-vertex model. The probability of a state is given by the
ratio of the weight of the state to the partition function of the
model
$$
P(state)=\frac{\prod_{\rm vertices} w(\rm vertex)}{Z}.
$$
This is the Gibbs measure of the $6$-vertex model.

Let us define the characteristic function of an edge $e$ as
\[
\sigma_e(state)=\left\{ \begin{array}{cc} 1,
&\mbox{if} \ e\,\,\mbox{is occupied by a path} ; \\
0, &  \mbox{otherwise}  \end{array} \right .
\]

A local correlation function is the expectation value of the
product of such characteristic functions:
$$
\langle\sigma_{e_1}\sigma_{e_2}..\sigma_{e_n}\rangle
=\sum_{states}P(state)\prod_{i=1}^n\sigma_{e_i}(state)
$$

\subsection{Boundary Conditions}

\subsubsection{} Let us fix arrows (or, equivalently, think edges) on outer
edges of $L_{N,M}$. States in the 6-vertex with the same
configurations of arrows on the boundary are called states  with
{\it fixed} boundary conditions. The difference between two such
states can occur only at inner edges.

The space of states with fixed boundary conditions is empty unless
the boundary values satisfy the ice rule: the total number of
incoming arrows on the boundary edges should be equal to the total
number of outgoing arrows. In the path formulation this means that
the number of paths through North and West boundaries should be
equal to the number paths through the South and East boundaries.

\begin{figure}[t]
\begin{center}
\includegraphics{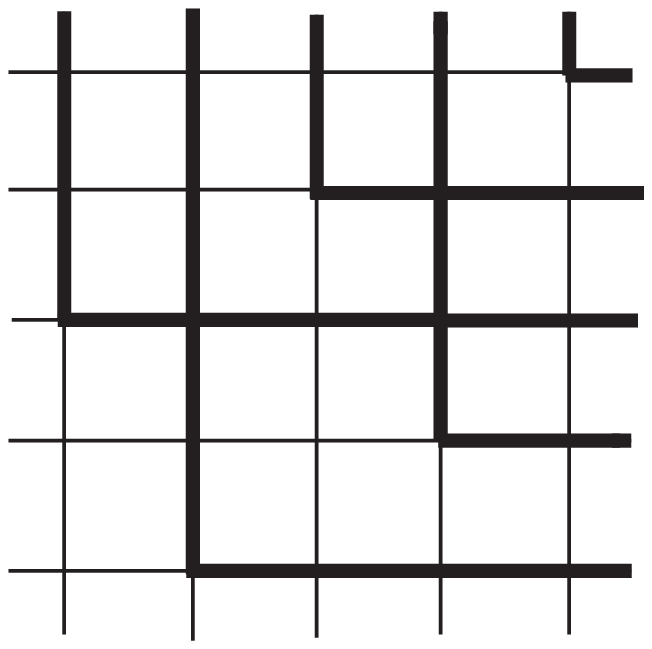}
\end{center}
\caption{A possible configuration of paths on a $5\times 5$ square
grid for the DW boundary conditions} \label{dwbc_conf}
\end{figure}

An example of such boundary conditions is the domain wall (DW)
boundary conditions. For the DW boundary conditions the arrows on
the boundary of the lattice are going into the lattice at the top
and bottom of the lattice and are going out of the lattice at the
right and left of it. A configuration of paths on a $5\times 5$
lattice with DW boundary conditions is presented on Fig.
\ref{dwbc_conf}.

Notice that  the differences
$$
n_a=n(a_1)-n(a_2),\quad n_b=n(b_1)-n(b_2),\quad n_c=n(c_2)-n(c_1)
$$
are the same for all configurations with given fixed boundary
conditions. Here $n(x)$ is the total number of vertices of type
$x=a_i,b_i,c_i$ in the configuration.

In particular, the partition function for fixed boundary
conditions trivially depends on magnetic fields:
\[
Z(a_1,a_2,b_1,b_2,c_1,c_2)=e^{(Hn_b+Vn_a)}
\left(\frac{c_2}{c_1}\right)^{\frac{n_c}{2}} Z(a,a,b,b,c,c)
\]

In this paper we will focus on the $6$-vertex model with fixed
boundary conditions.

\subsubsection{}
Another important type of boundary conditions are the {\it
periodic boundary conditions}. In this case the edges at opposite
sides of $L_{N,M}$ are identified so that the configuration of
arrows on the left and right boundary is the same as well as the
configuration of arrows on the top and bottom boundary.

The $6$-vertex model with periodic boundary conditions is an
example of an ``integrable'' (solvable) model in statistical
mechanics and has been studied extensively, see
\cite{Bax},\cite{LW} and references therein. In particular, it
means that the row-to-row transfer-matrix of the model can be
diagonalized by the Bethe ansatz.

\subsection{The Height Function}
By {\it outer faces} we mean unit squares centered at
$(-\frac{1}{2},m)$, $(N-\frac{1}{2},m)$, with
$m=-\frac{1}{2},\frac{1}{2},\dots, M-\frac{1}{2}$ and
$(n,-\frac{1}{2})$, $(n,N-\frac{1}{2})$, with
$n=-\frac{1}{2},\frac{1}{2},\dots, N-\frac{1}{2}$. Each corner
outer face has two edges in their boundary, other outer faces have
three edges in their boundary.

A height function $h$ is an integer-valued function on the faces
$F_N$ of the grid $L_N$ (including the outer faces), which is

\begin{itemize}

\item zero at the southwest corner of the grid,

\item non-decreasing when going up or right,

\item if $f_1$ and $f_2$ are neighboring faces, then
$|h(f_1)-h(f_2)|\leq 1$.

\end{itemize}

The {\it boundary value} of the height function is its restriction
to the ``outer faces''. Denote the set of outer faces by $\pa
F_N$. Given a function $h^{(0)}$ on $\pa F_N$ denote ${\mathcal
H}(h^{(0)})$ the space of all height functions with the boundary
value $h^{(0)}$.

If we enumerate faces by the coordinates of their centers the
height function can be regarded as $(N+1)\times (M+1)$ matrix with
non-negative entries.

It is clear that there is a bijection between states of the
$6$-vertex model with fixed boundary conditions and height
functions with fixed boundary values.

\begin{figure}[t]
\begin{center}
\includegraphics{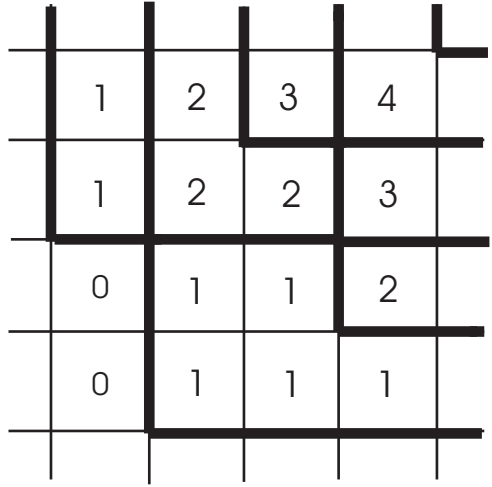}
\end{center}
\caption{Values of the height function for the configuration of
paths given on figure 2.} \label{hf_example}
\end{figure}

Indeed, given a height function consider its ``level curves'',
i.e. paths on the grid $L_N$, where the height function changes
its value by $1$, see Fig. \ref{hf_example}. Clearly, this defines
a state for the $6$-vertex model on $L_N$ with boundary conditions
determined by the boundary values of the height function.

On the other hand, given a state in the $6$-vertex model, consider
the corresponding configuration of paths. It is clear that there
is a unique height function whose level curves are these paths and
which satisfies the condition $h=0$ at the southwest corner.

It is clear that this correspondence is a bijection.

There is a natural partial order on the set of height functions
with given boundary values. One function is bigger then the other
if it is entirely  above the other. There exist the minimum
$h_{\rm min}$ and the maximum $h_{\rm max}$ height functions such
that $h_{\rm min}\leq h\leq h_{\rm max}$ for all height functions
$h$.

Thus, we can consider the $6$-vertex model as a theory of
fluctuating discrete surfaces constrained between $h_{\rm max}$
and $h_{\rm min}$. Each surface occurs with probability given by
the Boltzmann weights of the $6$-vertex model.

\subsection{The inhomogeneous $6$-vertex model and volume weights}

In the inhomogeneous $6$-vertex model the Boltzmann weights depend
on the edge. Thus, we have $6N^2$ parameters $a_i(m,n)$,
$b_i(m,n)$, and $c_i(m,n)$.

Let us assume that the inhomogeneity is only in magnetic fields,
i.e. weights $a$,$b$, and $c$ do not change from edge to edge, but
the magnetic fields $H(e)$, $V(e)$ do.

Let $\{P\}$ be the collection of paths corresponding to a state in
the $6$-vertex model and $\{h(f)\}$ be the corresponding height
function.

\begin{proposition} Let us assign weights $s(e)$ to the edges
of the lattice and $1$ to the outer edges, then
\[
\prod_{f}q_f^{h(f)}=\prod_{P}\prod_{e\in P} s(e)
\]
where $q_f=s(e_1)s(e_2)s(e_3)^{-1}s(e_4)^{-1}$ for inner faces,
$q_f=s(e)^{\e(e)}$ for the outer face $f$ adjacent to the edge
$e$, $q_f=1$ for outer faces not adjacent to any edge of the
lattice (corner faces) with $\e(e)=1$ for edges at the upper and
right sides of the boundary, $\e(e)=-1$ for edges at the lower and
left sides of the boundary.
\end{proposition}

The proof is an elementary exercise.

Let $s(e)=\exp(H(e)$ for horizontal edges and $s(e)=\exp(V(e))$
for vertical edges. Then the probability of the state with the
height function $h$ in such a model is
$$
P_q(state)=\frac{\prod_f q_f^{h(f)}\prod_{\rm vertices} w_0(\rm
vertex)}{Z}.
$$

If $H(n,m)=H+am$ and $V(n,m)=V+bn$, the weights $q_f$ are the same
for all faces $q_f=\exp(a+b)$ inside the lattice and the
probability is given by
$$
P(state)=\frac{q^{{\rm vol}(h)}\prod_{\rm vertices} w(\rm
vertex)}{Z},
$$
where $w(vertex)$ are the Boltzmann weights with constant magnetic
fields and
$$
{\rm vol}(h)=\sum_{f\in L_N} h(f)
$$
is the volume ``under'' the height function $h$.

\section{The thermodynamic limit}\label{f-energy}

\subsection{Stabilizing sequence of fixed boundary conditions}
\subsubsection{}Let $a=M/N$. We place the grid $L_{N,M}$  inside of the rectangle
$D=\{(x,y)|\,\,0\leq x \leq 1, 0\leq y\leq a\}$ so that the
vertices of the grid are the points with coordinates
$(\frac{n}{N+1},\frac{m}{N+1})$ where $n=1,\dots, N, \ m=1,\dots ,
M$.

We recall that a height function is a monotonic integer-valued
function on the faces of the grid, which satisfies Lipshitz
condition (it changes at most by $1$ on any two adjacent faces).
The height function can be regarded as a function on the centers
of the faces of the grid, i.e on $(n-1/2,m-1/2)$, where $n=0,\dots
,N+1, \ m=0,1,\dots, M+1$. The points $(\frac{1}{2},
m-\frac{1}{2})$, $(N+\frac{1}{2},m-\frac{1}{2})$,
$(n-\frac{1}{2},\frac{1}{2})$, and
$(n-\frac{1}{2},M+\frac{1}{2})$, where  $n=0,\dots, N+1, \ m=0,
\dots, M+1$ correspond to the ``outer'' faces of $L_{N,M}$.

We introduce the normalized height function as a piecewise linear
function on the unit square with the value
$$
h_N^{\rm norm}(x,y)=\frac{1}{N}h_N(n,m).
$$
for $\frac{n}{N+1}\leq x\leq \frac{n+1}{N+1}$ and
$\frac{m}{N+1}\leq y \leq \frac{m+1}{N+1}$. Here $h_N(n,m)$ is a
height function on $L_{N,M}$. Normalized height functions are
nondecreasing in $x$ and $y$ directions and they  satisfy:
\begin{equation}\label{Lip}
h(x,y)-h(x',y')\leq x-x'+y-y'.
\end{equation}
if $x\geq x'$ and $y\geq y'$.

The boundary value of the normalized height function defines a
piece-wise constant monotonic function on each side of the region
which changes by $\pm 1/N$ or do not change between two
neighboring boundary sites.

Denote the space of such normalized height functions with the
boundary value $h_0$ by $L_{N,M}( h_0)$.

There is a natural partial ordering on the set of all normalized
height functions with given boundary values descending from the
partial order on height functions: $h_1\geq h_2$ if $h_1(x)\geq
h_2(x)$ for all $x\in D$. We define the operations
\[
h_1\vee h_2=min_{x\in D} (h_1(x),h_2(x)), \ h_1\wedge
h_2=max_{x\in D} (h_1(x),h_2(x)) ,
\]
It is clear that
\[
h_1\wedge h_2\geq h_1,h_2 \geq h_1\vee h_2
\]
It is also clear that in this partial order there is unique
minimal and unique maximal height functions, which we denote by
$h_N^{min}$ and $h_N^{max}$, respectively.

A {\it sequence of stabilizing fixed boundary conditions} is a
sequence of functions $h^{(N)}_0$ which are boundary values of a
normalized height function, and  which converges
as $N\to \infty$ to a function $\phi_0$
on the boundary of $D$ which is  non-decreasing along $x$ and $y$
direction, $\phi(0,0)=0$, and satisfies the condition:

\begin{equation}\label{Lip-cont}
\phi(x,y)-\phi(x',y')\leq x-x'+y-y'
\end{equation}

We will call the function $\phi_0$ the boundary condition for the
domain $D$ (or simply the boundary condition). Any function on $D$
which satisfies (\ref{Lip-cont}), which is non-decreasing in $x$
and $y$ directions, and coinside with $\phi_0$ at the boundary is
called a {\emph height function} on $D$ with boundary condition
$\phi_0$.

Among all possible boundary conditions $\phi_0$ we will
distinguish piecewise linear boundary conditions with the slope
$0$ or $1$ along the coordinate axes. We will call these boundary
values {\bf critical}. It is clear that any boundary conditions
can be approximated critical boundary conditions.

\subsection{The thermodynamic limit}
If $q$ is fixed and is not equal to $1$ in the large volume limit,
the system will be in a neighborhood of the minimal height
function for $q<1$ and in the neighborhood of the maximal height
function for $q>1$. One should expect that the partition function
and local correlation functions will have finite limit.

When $q=\exp(\frac{\lambda}{N})$ for some $\lambda$, one should
expect the existence of the limit shape. We will study this limit
in Section \ref{section_var}.

\subsection{Gibbs measures with fixed slope}

\begin{definition} The Gibbs measure of the $6$-vertex model on an infinite
lattice has the {\it slope} $(h,v)$ if
\[
\lim_{k\to\infty}<\frac{h(n+k,m)-h(n,m)}{k}>=h
\]
and
\[
\lim_{k\to\infty}<\frac{h(n,m+k)-h(n,m)}{k}>=v
\]
\end{definition}

It is clear from the definition of the height function that the
slope should satisfy conditions $0\leq h,v \leq 1$.

The slope is simply the average number of horizontal and vertical
edges occupied by paths per length.

The important corollary of the result of \cite{Shef} is that, when
a gradient Gibbs measure satisfies certain convexity conditions,
there exists a unique translationally invariant measure. This
implies that for the $6$-vertex model one should expect the
uniqueness of such a measure for the generic slope.

Translationally invariant measures can be obtained by taking the
thermodynamic limit of the 6-vertex model with magnetic fields on
a torus. Then the slope is the Legendre conjugate to magnetic
fields.
$$
h=\lim_{N,M\to \infty}\langle\sigma_{e}\rangle=\lim_{N,M\to
\infty}\langle \frac{N(hor)}{N}\rangle=\lim_{N,M\to
\infty}\frac{1}{2NM}\frac{\partial\log Z}{\partial H}+\frac{1}{2}
$$
for a horizontal $e$ and
$$
v=\lim_{N.M\to \infty}\langle\sigma_{e}\rangle=\lim_{N,M\to
\infty}\langle \frac{N(ver)}{N}\rangle=\lim_{N,M\to
\infty}\frac{1}{2NM}\frac{\partial\log Z}{\partial V}+\frac{1}{2}
$$
for a vertical $e$.

\section{The thermodynamic limit of the 6-vertex model for
the periodic boundary conditions}

The free energy per site in the thermodynamic limit is
\[
f=-\lim_{N,M\to\infty}\frac{\log(Z_{N,M})}{NM}
\]
where $Z_{N,M}$ is the partition function with the periodic
boundary conditions on the rectangular grid $L_{N,M}$. It is a
function of the Boltzmann weights and magnetic fields.

\begin{remark} Normally, it is expected that the free energy is not
identically zero. Physically, this means that the ``excitations''
have the characteristic length which is much smaller than the
characteristic length of the system. In some cases the free energy
is identically zero, then one expects that a typical excitation
will be comparable with the size of the system.
\end{remark}

For generic $H$ and $V$ the $6$-vertex model in the thermodynamic
limit has the translationally invariant Gibbs measure with the
slope $(h,v)$:
\begin{equation}\label{slope}
h=-\frac{1}{2}\frac{\pa f}{\pa H}+\frac{1}{2}, \quad
v=-\frac{1}{2}\frac{\pa f}{\pa V}+\frac{1}{2}.
\end{equation}

The parameter
$$
\Delta=\frac{a^2+b^2-c^2}{2ab}.
$$
defines many characteristics of the $6$-vertex model in the
thermodynamic limit.

\subsection{The phase diagram for $\Delta>1$} The weights $a$, $b$, and $c$
in this region satisfy one of the two inequalities, either $a>b+c$
or $b>a+c$.

If $a>b+c$, the Boltzmann weights $a$, $b$, and $c$ can be
parameterized as
\begin{equation}\label{dg1par-1}
a=r\sinh(\lambda+\eta), b=r\sinh(\lambda), c=r\sinh(\eta)
\end{equation}
with $\lambda,\eta>0$.

If $a+c<b$, the Boltzmann weights can be parameterized as
\begin{equation}\label{dg1par-2}
a=r\sinh(\lambda-\eta), b=r\sinh(\lambda), c=r\sinh(\eta)
\end{equation}
with $0<\eta<\lambda$.

For both of these parametrization of weights $\Delta=\cosh(\eta)$.

\begin{figure}[t]
\begin{center}
\includegraphics[width=8cm, height=8cm]{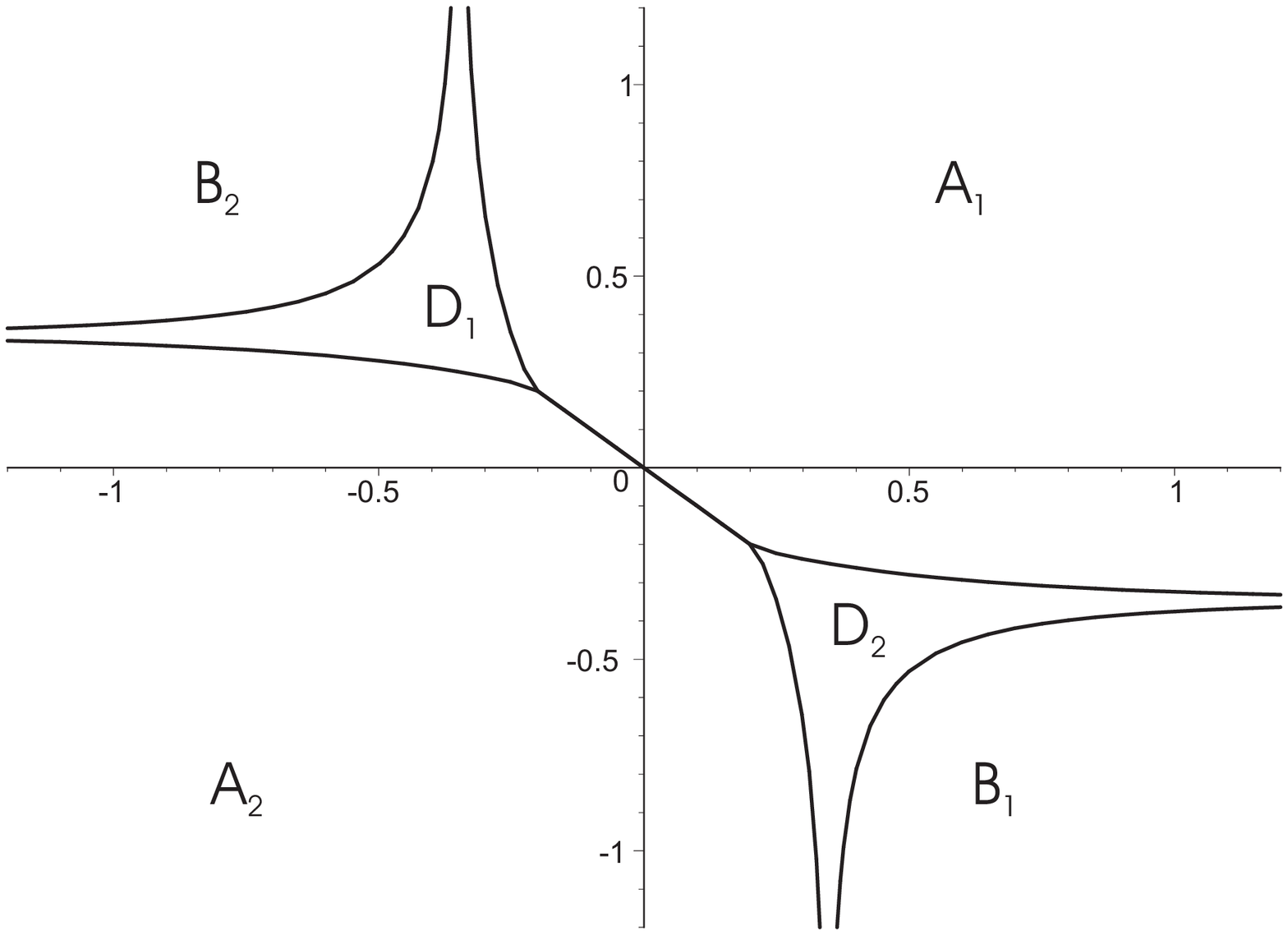}
\end{center}
\caption{The phase diagram in the $(H,V)$-plane for $a=2$, $b=1$,
and $c=0.8$} \label{ferro_diagram_1}
\end{figure}

\begin{figure}[t]
\begin{center}
\includegraphics[width=8cm, height=8cm]{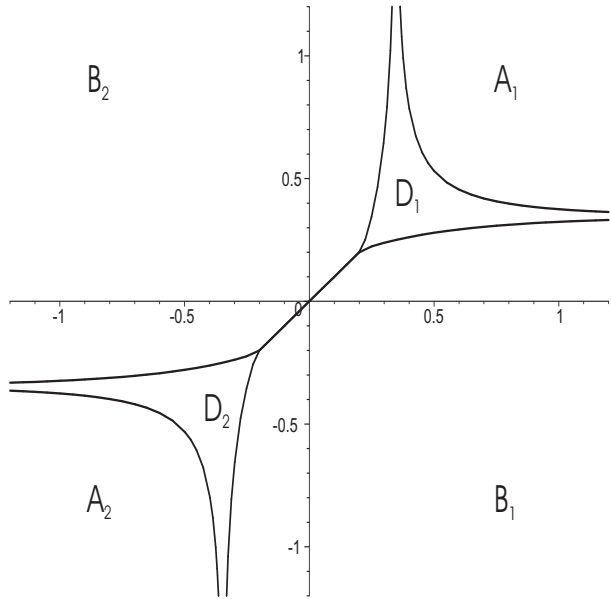}
\end{center}
\caption{The phase diagram in the $(H,V)$-plane for $a=1$, $b=2$,
and $c=0.8$} \label{ferro_diagram_2}
\end{figure}

The phase diagram of the model for $a>b+c$ (and, therefore, $a>b$)
is shown on Fig. \ref{ferro_diagram_1} and for $b>a+c$ (and,
therefore, $a<b$) on Fig. \ref{ferro_diagram_2}.

\begin{figure}[b]
\begin{center}
\includegraphics{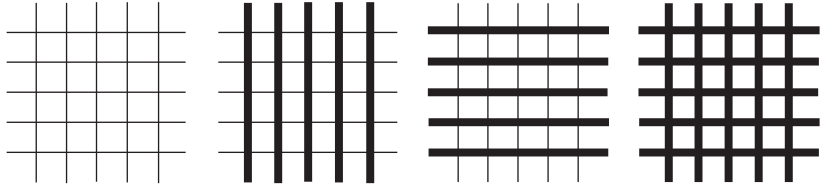}
\end{center}
\caption{Four frozen configurations of the ferromagnetic phase}
\label{ferroelectric_phase}
\end{figure}

When magnetic fields $(H,V)$ are in one of the regions $A_i, B_i$
of the phase diagram, the system in the thermodynamic limit has
the translationally invariant Gibbs measure supported on the
corresponding frozen configuration. There are four frozen
configurations $A_1$, $A_2$, $B_1$, and $B_2$, shown on Fig.
\ref{ferroelectric_phase}. For a finite but large grid the
probability of any other state is at most of order $\exp(-\alpha
N)$ for some positive $\alpha$.

Local correlation functions are given by the value of the
corresponding observable on the frozen state
$$
\lim_{N\to\infty}\langle \sigma_{e_1}\dots\sigma_{e_n}\rangle_N=
\sigma_{e_1}(S)\dots\sigma_{e_n}(S)
$$
where $S$ is the one of the ferromagnetic states $A_i,B_i$.

The frozen regions in the $(H,V)$-plane are described by the set
of inequalities. The boundaries of these regions can be derived by
analyzing the next to the largest eigenvalue of the row-to-row
transfer matrix. The description is separated into two cases:
$a>b+c$ and $b>a+c$. Notice that $a\neq b$ since $\Delta>1$.

$\bullet$ $a>b+c$, see Fig. \ref{ferro_diagram_1},
\begin{eqnarray}
&& \mbox{$A_1$-region:}\qquad V+H\geq 0, \qquad
\cosh(2H)\leq\Delta, \nonumber
\\
&& \qquad (e^{2H}-b/a)(e^{2V}-b/a)\geq(c/a)^2,\qquad e^{2H}>b/a,
\qquad \cosh(2H)>\Delta, \nonumber
\\
&& \mbox{$A_2$-region:}\qquad V+H\leq 0, \qquad
\cosh(2H)\leq\Delta, \nonumber
\\
&& \qquad (e^{-2H}-b/a)(e^{-2V}-b/a)\geq(c/a)^2,\qquad
e^{-2H}>b/a, \qquad \cosh(2H)>\Delta, \nonumber
\\
&& \mbox{$B_1$-region:}\qquad
(e^{2H}-a/b)(e^{-2V}-a/b)\geq(c/b)^2, \qquad e^{2H}>a/b, \nonumber
\\
&& \mbox{$B_2$-region:}\qquad
(e^{-2H}-a/b)(e^{2V}-a/b)\geq(c/b)^2, \qquad e^{-2H}>a/b.
\nonumber
\end{eqnarray}

$\bullet$ $b>a+c$, see Fig. \ref{ferro_diagram_2},
\begin{eqnarray}
&& \mbox{$A_1$-region:}\qquad(e^{2H}-b/a)(e^{2V}-b/a)\geq(c/a)^2,
\qquad e^{2H}> b/a; \nonumber
\\
&&
\mbox{$A_2$-region:}\qquad(e^{-2H}-b/a)(e^{-2V}-b/a)\geq(c/a)^2,
\qquad e^{-2H}> b/a; \nonumber
\\
&& \mbox{$B_1$-region:}\qquad V-H\geq 0, \qquad
\cosh(2H)\leq\Delta, \nonumber
\\
&& \qquad (e^{2H}-a/b)(e^{-2V}-a/b)\geq(c/b)^2, \qquad e^{2H}>a/b,
\qquad \cosh(2H)>\Delta, \nonumber
\\
&& \mbox{$B_2$-region:}\qquad  V-H\leq 0, \qquad
\cosh(2H)\leq\Delta, \nonumber
\\
&& \qquad (e^{-2H}-a/b)(e^{2V}-a/b)\geq(c/b)^2, \qquad
e^{-2H}>a/b, \qquad \cosh(2H)>\Delta. \nonumber
\end{eqnarray}

The free energy is a linear function in $H$ and $V$ in the four
frozen regions:
\begin{eqnarray}
&& f=-\ln a-H-V\qquad\mbox{in}\quad A_1, \nonumber
\\
&& f=-\ln b+H-V\qquad\mbox{in}\quad B_2,
\\
&& f=-\ln a+H+V\qquad\mbox{in}\quad A_2, \nonumber
\\
&& f=-\ln b-H+V\qquad\mbox{in}\quad B_1. \nonumber
\label{linear_energy}
\end{eqnarray}

If $(H,V)$ is a region $D_1$ or $D_2$  the corresponding
translationally invariant Gibbs measure has the slope $(h,v)$
given by (\ref{slope}). In this phase the system is disordered,
which means that local correlation functions decay as a power of
the distance $d(e_i,e_j)$ between $e_i$ and $e_j$ when
$d(e_i,e_j)\to \infty$.

In the regions $D_1$ and $D_2$ the free energy is given by
\cite{SY}:
\begin{eqnarray}
f(H,V)&=&\min\Bigl\{\min_\alpha\Bigl\{E_1-H-(1-2\alpha)V-\frac{1}{2\pi
i} \int_C\ln(\frac{b}{a}-\frac{c^2}{ab-a^2 z})\rho(z)dz\Bigr\},
\nonumber
\\
&& \min_\alpha\Bigl\{E_2+H-(1-2\alpha)V-\frac{1}{2\pi i}
\int_C\ln(\frac{a^2-c^2}{ab}+\frac{c^2}{ab-a^2
z})\rho(z)dz\Bigr\}\Bigr\}, \label{free_energy}
\end{eqnarray}
where $\rho(z)$ can be found from the integral equation
\begin{equation}\label{int-eqn}
\rho(z)=\frac{1}{z}+\frac{1}{2\pi i}\int_C
\frac{\rho(w)}{z-z_2(w)}dw -\frac{1}{2\pi i}\int_C
\frac{\rho(w)}{z-z_1(w)}dw,
\end{equation}
in which
$$
z_1(w)=\frac{1}{2\Delta-w}, \qquad z_2(w)=-\frac{1}{w}+2\Delta.
$$
$\rho(z)$ satisfies the following normalization condition:
$$
\alpha=\frac{1}{2\pi i}\int_C\rho(z)dz.
$$
The contour of integration $C$ (in the complex $z$-plane) is
symmetric with respect to the conjugation $z\rightarrow\bar{z}$,
is dependent on $H$ (see Appendix \ref{a_free_energy_hom}) and is
defined by the condition that the form $\rho(z)dz$ has purely
imaginary values on the vectors tangent to $C$:
$$
{\rm Re}(\rho(z)dz)\Bigr|_{z\in\,\, C}=0.
$$

The formula (\ref{free_energy}) for the free energy follows from
the Bethe Ansatz diagonalization of the row-to-row
transfer-matrix. Its derivation is outlined in Appendix
\ref{a_free_energy_hom}. It relies on a number of conjectures that
are supported by numerical and analytical evidence and in physics
are taken for granted. However, there is no rigorous proof.

There are two points where three phases coexist (two frozen and
one disordered phase). These points are called {\it tricritical}.
The angle $\theta$ between the boundaries of $D_1$ (or $D_2$) at a
tricritical point is given by
$$
\cos(\theta)=\frac{c^2}{c^2+2\min(a,b)^2(\Delta^2-1)}.
$$
The existence of such points makes the $6$-vertex model (and its
degeneration known as the $5$-vertex model \cite{HWKK}) remarkably
different from dimer models \cite{KOS} where generic singularities
in the phase diagram are cusps. Physically, the existence of
singular points where two curves meet at the finite angle
manifests the presence of interaction in the $6$-vertex model.

Notice that when $\Delta= 1$ the phase diagram of the model has a
cusp at the point $H=V=0$. This is the transitional point between
the region $\Delta>1$ and the region $|\Delta|<1$ which is
described below.

\subsection{The phase diagram $|\Delta|<1$} In this case,
the Boltzmann weights have a convenient parameterization by
trigonometric functions. When $1\geq\Delta\geq 1$
\[
a=r\sin(\lambda-\gamma), b=r\sin(\lambda), c=r\sin(\gamma),
\]
where $0\leq \gamma\leq \pi/2$, $\gamma\leq \lambda\leq \pi$, and
$\Delta=\cos\gamma$.

When $0\geq\Delta\geq -1$
\[
a=r\sin(\gamma-\lambda), b=r\sin(\lambda), c=r\sin(\gamma),
\]
where $0\leq \gamma\leq \pi/2$, $\pi- \gamma\leq \lambda\leq \pi$,
and $\Delta=-\cos\gamma$.

\begin{figure}[t]
\begin{center}
\includegraphics[width=10cm, height=10cm]{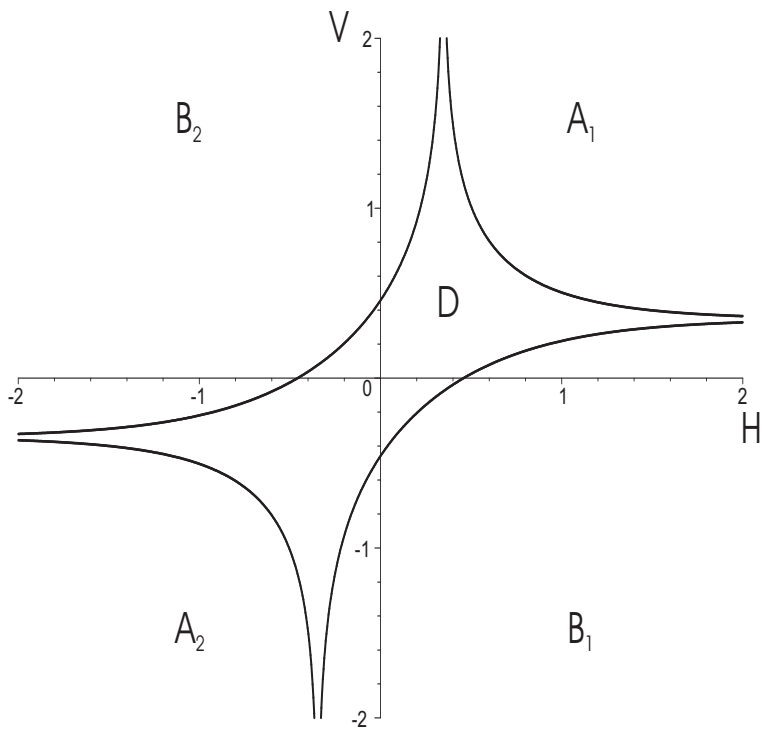}
\end{center}
\caption{The phase diagram in the $(H,V)$-plane for $a=1$, $b=2$,
and $c=2$} \label{disordered_diagram}
\end{figure}

The phase diagram of the $6$-vertex model with $|\Delta|<1$ is
shown on Fig. \ref{disordered_diagram}. The phases $A_i, B_i$ are
frozen and identical to the frozen phases for $\Delta>1$. The
phase $D$ is disordered. For magnetic fields $(H,V)$ the Gibbs
measure is translationally invariant with the slope
$(h,v)=(\frac{\pa f(H,V)}{\pa H}, \frac{\pa f(H,V)}{\pa V})$.

The frozen phases can be described by the following inequalities:
\begin{eqnarray}
&& \mbox{$A_1$-region:}\qquad(e^{2H}-b/a)(e^{2V}-b/a)\geq(c/a)^2,
\qquad e^{2H}> b/a, \nonumber
\\
&&
\mbox{$A_2$-region:}\qquad(e^{-2H}-b/a)(e^{-2V}-b/a)\geq(c/a)^2,
\qquad e^{-2H}> b/a, \nonumber
\\
&& \mbox{$B_1$-region:}\qquad(e^{2H}-a/b)(e^{-2V}-a/b)\geq(c/b)^2,
\qquad e^{2H}>a/b, \label{boundaries_phases}
\\
&& \mbox{$B_2$-region:}\qquad(e^{-2H}-a/b)(e^{2V}-a/b)\geq(c/b)^2,
\qquad e^{-2H}>a/b. \nonumber
\end{eqnarray}

The free energy function in the frozen regions is still given by
the formulae (\ref{linear_energy}). The first derivatives of the
free energy are continuous at the boundary of frozen phases, The
second derivative is continuous in the tangent direction at the
boundary of frozen phases and is singular in the normal direction.

It is smooth  in the disordered region where it is given by
(\ref{free_energy}) which, as in case $\Delta>1$ involves a
solution to the integral equation (\ref{int-eqn}). The contour of
integration in (\ref{int-eqn}) is closed for zero magnetic fields
and, therefore, the equation (\ref{int-eqn}) can be solved
explicitly by the Fourier transformation \cite{Bax} .

The $6$-vertex Gibbs measure with zero magnetic fields converges
in the thermodynamic limit to the superposition of translationally
invariant Gibbs measures with the slope $(1/2,1/2)$. There are two
such measures. They correspond to the double degeneracy of the
largest eigenvalue of the row-to-row transfer-matrix \cite{Bax}.

There is a very interesting relationship between the $6$-vertex
model in zero magnetic fields and the highest weight
representation theory of the corresponding quantum affine algebra.
The double degeneracy of the Gibbs measure with the slope
$(1/2,1/2)$ corresponds to the fact that there are two integrable
irreducible representations of $\widehat{sl_1}$ at level one.
Correlation functions in this case can be computed using
$q$-vertex operators \cite{JM}. For latest developments see
\cite{BJMST}.

\subsection{The phase diagram $\Delta<-1$}

\subsubsection{The phase diagram}The Boltzmann weights for these values of
$\Delta$ can be conveniently parameterized as
\begin{equation}\label{hyp-param}
a=r\sinh(\eta-\lambda), b=r\sinh(\lambda), c=r\sinh(\eta),
\end{equation}
where $0<\lambda<\eta$ and $\Delta=-\cosh\eta$.

\begin{figure}[t]
\begin{center}
\includegraphics[height=10cm, width=10cm]{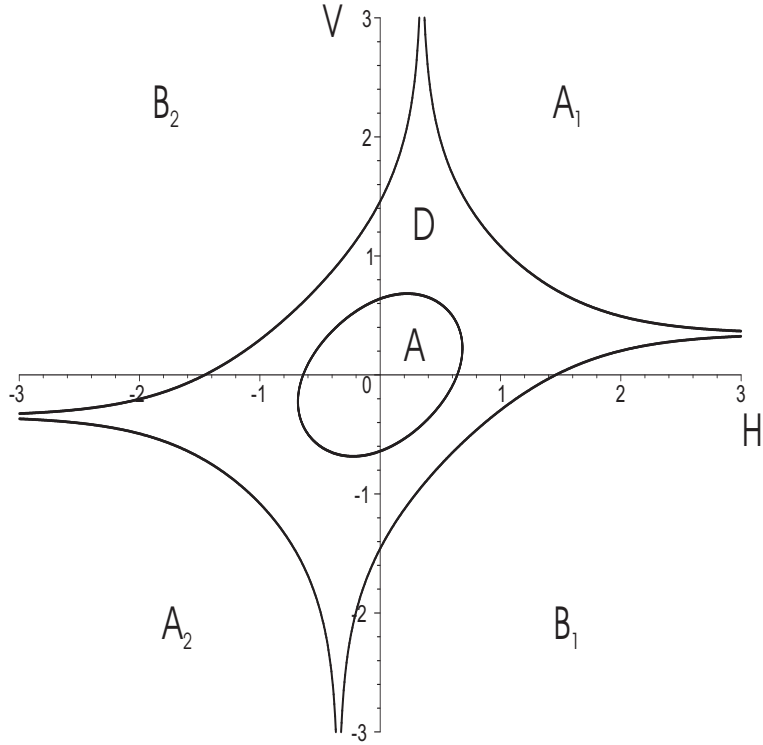}
\end{center}
\caption{The phase diagram in the $(H,V)$-plane for $a=1$, $b=2$,
and $c=6$} \label{antiferroelectric_diagram}
\end{figure}

The Gibbs measure in thermodynamic limit depends on the value of
magnetic fields. The phase diagram in this case is shown on Fig.
\ref{antiferroelectric_diagram} for $b/a>1$. In the
parameterization (\ref{hyp-param}) this correspond to $0<\lambda<
\eta/2$. When $\eta/2<\lambda <\eta$ the $4$-tentacled ``amoeba''
is tilted in the opposite direction as on Fig.
\ref{ferro_diagram_1}. When $(H,V)$ is in one of the $A_i,B_i$
regions in the phase diagram the Gibbs measure is supported on the
corresponding frozen configuration, see Fig.
\ref{ferroelectric_phase}.

The $A_i,B_i$ regions on the phase diagram are defined by
inequalities (\ref{boundaries_phases}). The free energy in these
regions is linear and is given by (\ref{linear_energy}).

If $(H,V)$ is in the region $D$, the Gibbs measure is the
translationally invariant measure with the slope $(h,v)$
determined by (\ref{slope}). The free energy in this case is
determined by solutions to the linear integral equation
(\ref{int-eqn}) and is given by the formula (\ref{free_energy}).

\begin{figure}[b]
\begin{center}
\includegraphics[width=8cm, height=4cm]{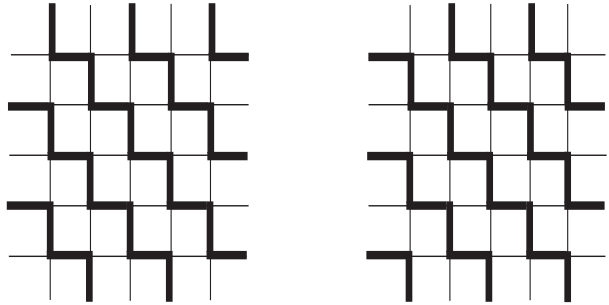}
\end{center}
\caption{Two most probable configurations in the antiferromagnetic
phase.} \label{antiferr}
\end{figure}

If $(H,V)$ is in the region $A$, the Gibbs measure is the
superposition of two Gibbs measures with the slope $(1/2,1/2)$. In
the limit $\Delta\to -\infty$ these two measures degenerate to two
measures supported on configurations $C_1, C_2$, respectively,
shown on Fig. \ref{antiferr}. For a finite $\Delta$ the support of
these measures consists of configurations which differ from $C_1$
and $C_2$ in finitely many places on the lattice.

\begin{figure}[h]
\begin{center}
\includegraphics{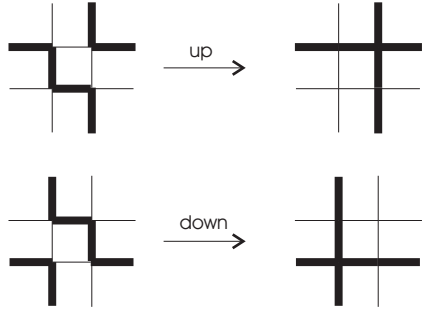}
\end{center}
\caption{The elementary up and down fluctuations in the
antiferromagnetic phase.} \label{fluct}
\end{figure}

We notice that any two configurations lying in the support of each
of these Gibbs measures can be obtained from $C_1$ or $C_2$ via
flipping the path at a vertex ``up'' or ``down'' as it is shown on
Fig. \ref{fluct} finitely many times. It is also clear that it
takes infinitely many flips to go from $C_1$ to $C_2$.

The $6$-vertex model in the phase $A$ is disordered and is also
noncritical.

Here non-criticality means that the local correlation function
$\langle\sigma_{e_i}\sigma_{e_j}\rangle$ decays as $\exp(-\alpha
d(e_i,e_j))$ with some positive $\alpha$ as the distance
$d(e_i,e_j)$ between $e_i$ and $e_j$ increases to infinity.

The free energy in the $A$-region can be explicitly computed by
solving the equation (\ref{int-eqn}) via the Fourier transform
\cite{Bax}.

\subsubsection{The antiferromagnetic region}\label{AF-region}
This boundary of the antiferromagnetic region $A$ can be derived
similarly to the boundaries of the ferromagnetic regions $A_i$ and
$B_i$  by analyzing next to the largest eigenvalue of the
row-to-row transfer-matrix. The difference is that for the region
$A$ the largest eigenvalue will correspond to $n=N/2$ and to
compute it we should use the solution to the integral equation
(\ref{int-eqn}) in the case when the contour of integration is
closed.

This computation was done in \cite{SY}, \cite{LW}. The result is a
simple closed curve, which can be described parameterically as
$$
H(s)=\Xi(s), \qquad V(s)=\Xi(\eta-\theta_0+s),
$$
where
$$
\Xi(\varphi)=\cosh^{-1}\Bigl(\frac{1}{{\rm
dn}(\frac{K}{\pi}\varphi|1-\nu)}\Bigr),
$$
$$
|s|\leq 2\eta,
$$
and
$$
e^{\theta_0}=\frac{1+\max(b/a,a/b)e^{\eta}}{\max(b/a,a/b)+e^{\eta}}.
$$
The parameter  $\nu$ is defined by the equation $\eta K(\nu)=\pi
K'(\nu)$, where
$$
K(\nu)=\int_0^{\pi/2}(1-\nu\sin^2(\theta))^{-1/2}d\theta \qquad
K'(\nu)=\int_0^{\pi/2}(1-(1-\nu)\sin^2(\theta))^{-1/2}d\theta.
$$

The curve is invariant with respect to the reflections $(H,V)\to
(-H,-V)$ and $(H,V)\to (V,H)$ since the function $\Xi$ satisfies
the identities
$$
\Xi(\varphi)=-\Xi(-\varphi), \qquad
\Xi(\eta-\varphi)=\Xi(\eta+\varphi).
$$
This function is also $4\eta$-periodic:
$\Xi(4\eta+\varphi)=\Xi(\varphi)$.

\begin{proposition}
The boundary of the antiferromagnetic region is a real algebraic
curve in $e^H$ and $e^V$ given by
\begin{eqnarray}
&& \Bigl((1-\nu\cosh^2V_0) \cosh^2H+\sinh^2V_0-(1-\nu)\cosh V_0
\cosh H \cosh V\Bigr)^2= \nonumber
\\
&& (1-\nu\cosh^2V_0)\sinh^2V_0\cosh^2V\sinh^2H(1-\nu\cosh^2H),
\label{antiferroelectic_boundary}
\end{eqnarray}
where $V_0$ is the positive value of $V$ on the curve when $H=0$.
Notice that $\nu$ depends on the Boltzmann weights $a,b,c$ only
through $\eta$.
\end{proposition}

\begin{proof} The parametric description of the boundary curve
implies that
$$
{\rm dn}(\frac{K}{\pi}s|1-\nu)=\frac{1}{\cosh H}\qquad {\rm
dn}(\frac{K}{\pi}(\eta -\theta_0+s)|1-\nu)=\frac{1}{\cosh V}
$$
The addition formula for the Jacobi elliptic function $\dn$
\cite{AS}
$$
\dn(u+v)=\frac{\dn u\dn v-(1-m)\sn u\cn u\sn v\cn
v}{1-(1-m)\sn^2u\sn^2v}.
$$
can be used to express $\dn u$ and $\dn v$ for
$u=K(\eta-\theta_0)/\pi$ and $v=K s/\pi$ in terms of $N$ and $V$.
Using elementary identities for the elliptic functions, we obtain
$$
\sn^2u\cn^2u(\dn^2v-\nu)(1-\dn^2v)=\bigl(\dn u\dn v
-\dn(u+v)(\cn^2u+\sn^2u\dn^2v)\bigr)^2
$$
For $u=K(\eta-\theta_0)/\pi$ and $v=K s/\pi$ this identity turns
into
$$
\Bigl(\cn^2u \cosh^2H+\sn^2u-\dn u \cosh H \cosh V\Bigr)^2=
\cn^2u\sn^2u\cosh^2V(\cosh^2H-1)(1-m\cosh^2H).
$$
Denote ${\rm dn}(\frac{K}{\pi}(\eta -\theta_0)|1-m)=1/\cosh V_0$,
then this identity becomes (\ref{antiferroelectic_boundary}).
\end{proof}

\section{Some asymptotics of the free energy}

\subsection{The scaling near the boundary of the $D$-region}

Here we study the asymptotics of the free energy of the $6$-vertex
model as the point $(H,V)$ inside a disordered region approaches
its boundary on the phase diagram of the model.

Let us consider the interface between the disordered region $D$
($D_2$ for $\Delta>1$) and the $A_1$-region, see Fig.
\ref{disordered_diagram}. It is given by $g(H,V)=0$, where
\begin{equation}\label{inter_D_F}
g(H,V)=\ln(b/a+\frac{c^2/a^2}{e^{2H}-b/a})-2V.
\end{equation}

Let $\vec{H}_0=(H_0,V_0)$ be a regular point on the interface,
i.e. the interface can be parameterized by real analytic functions
in its neighborhood, then $g(H_0,V_0)=0$.

We denote the normal vector to the interface at $\vec{H}_0$ by
$\vec{n}$ and the tangent vector by $\vec{\tau}$. A point
$\vec{H}=(H,V)$ in the vicinity of $\vec{H_0}$ can be represented
as $\vec{H}(r,s,t)=\vec{H_0}+r^2s\vec{n}+rt\vec{\tau}$, where $s$
and $t$ are local coordinates in the normal and tangent
directions, respectively, and $r$ is a scaling factor such that
$r\rightarrow 0$. Let us choose the normal vector $\vec{n}$ so
that it points in the direction of the disordered region $D$, then
the point $\vec{H}(r,s,t)$ belongs to $D$ if $s\geq 0$ .

\begin{theorem}
Let $\vec{H}(r,s,t)$ be defined as above. The asymptotics of the
free energy of the $6$-vertex model in the limit $r\rightarrow 0$
is given by
\begin{equation}\label{asympt_free_energy_interface}
f(\vec{H}(r,s,t))=f_{lin}(\vec{H}(r,s,t)) +\eta(s,t)r^3+O(r^5),
\end{equation}
where $f_{lin}(H,V)=-\ln(a)-H-V$ and
\begin{equation}\label{scal_lim_bound}
\eta(s,t)=-\kappa \left(\theta s+t^2\right)^{3/2}.
\end{equation}
Here the constants $\kappa$ and $\theta$ depend on the Boltzmann
weights of the model and on $(H_0,V_0)$ and are given by
$$
\kappa=\frac{16}{3\pi}\,\partial_H^2 g(H_0,V_0)
$$
and
$$
\theta=\frac{4+\left(\partial_H g(H_0,V_0)\right)^2}{2\partial_H^2
g(H_0,V_0)},
$$
where $g(H,V)$ is defined in (\ref{inter_D_F}).

Moreover, $\partial_H^2g(H_0,V_0)>0$ and, therefore, $\theta>0$.
\end{theorem}

We refer the reader to \cite{P} for the details.

\subsection{The scaling in the tentacle}
Assume that $a>b$. The theorem below describes the asymptotic of
the free energy function when $H\to +\infty$ and
\begin{equation}
\frac{1}{2}\ln(b/a)-\frac{c^2}{2ab}\,e^{-2H}\leq V
\leq\frac{1}{2}\ln(b/a)+\frac{c^2}{2ab}\,e^{-2H}, \qquad
H\longrightarrow\infty. \label{tentacle}
\end{equation}
These values of $(H,V)$ describe points inside the right
``tentacle'' on the Fig. \ref{ferro_diagram_1}.

Let us parameterize these values of $V$ as
$$
V=\frac{1}{2}\ln(b/a)+\beta\frac{c^2}{2ab}e^{-2H},
$$
where $\beta\in[-1,1]$.

\begin{theorem}
When $H\to \infty$ and $\beta\in [-1,1]$ the asymptotic of the
free energy is given by the following formula:
\begin{eqnarray}\label{tent-ass}
f(H,V)=-\frac{1}{2}\ln(ab)-H-\nonumber
\\  \frac{c^2}{2ab}\,e^{-2H} \Bigl(\beta+
&&
\frac{2}{\pi}\sqrt{1-\beta^2}-\frac{2}{\pi}\beta\arccos(\beta)\Bigr)
+O(e^{-4H}), \nonumber
\end{eqnarray}
\end{theorem}
\begin{proof}
From the integral equation for $\rho(z)$ we can derive the large
$H$ asymptotics of the density function:
\begin{equation}\label{tent-dens}
\rho(z)=\frac{1}{z}+\frac{2\Delta\alpha}{z^2}+..., \qquad
|z|\rightarrow\infty
\end{equation}
The integration contour is symmetric with respect to complex
conjugation $z\to \bar{z}$. The contour is a small deformation of
the segment of the circle of radius $e^{2H}$ centered at the
origin with endpoints having arguments $\pm \pi\alpha$.

For large $H$ the free energy function is given by
\begin{eqnarray}
f(H,V)&=&\min_{0\leq\alpha\leq 1}\Bigl(-\ln a-H-(1-2\alpha)V
\nonumber
\\
&& -\frac{1}{2\pi i}\int_C\ln(\frac{b}{a}
+\frac{c^2}{a^2z-ab})\rho(z)dz\Bigr). \nonumber
\end{eqnarray}

As $H\to \infty$ the density is given by (\ref{tent-dens}) and,
taking into account the asymptotical description of the contour of
integration we obtain
\begin{eqnarray}
f(H,V)&=&\min_{\alpha}\Bigl(-\ln a -H-V+\alpha(2V-\ln(b/a))
\nonumber
\\
&& -\frac{c^2}{ab}\,e^{-2H}\kappa(\alpha)+O(e^{-4H})\Bigr),
\nonumber
\end{eqnarray}
where
$$
e^{-2H}\kappa(\alpha)=\int_{C}\frac{dw}{w^2}
$$
This integral is easy to compute:
$$
\kappa(\alpha)=\frac{\sin(\pi\alpha)}{\pi}.
$$
The minimum occur at
\begin{equation}
\cos(\pi\alpha)=\frac{ab}{c^2}\bigl(2V-\ln(b/a)\bigr)e^{2H}.
\label{polar_tent}
\end{equation}
or, at
\[
\cos(\pi \alpha)=\beta
\]

The formula (\ref{tent-ass}) follows after the substitution of
this into the expression for $f(H,V)$.
\end{proof}

\subsection{The $5$-vertex limit}
The $5$-vertex model can be obtained as the limit of the
$6$-vertex model when $\Delta\rightarrow\infty$. Magnetic fields
in this limit behave as follows:
\begin{itemize}

\item $a>b+c$. In the parameterization (\ref{dg1par-1}) after
changing variables $H=\frac{\eta}{2}+l$, and $V=-\frac{\eta}{2}+m$
take the limit $\eta\to \infty$ keeping $\lambda$ fixed. The
weights will converge (up to a common factor) to:
\[
a_1:a_2:b_1:b_2:c_1:c_2\to e^{\lambda+l+m}:e^{\lambda-l-m}:
(e^\lambda-e^{-\lambda})e^{l-m}:0:1:1
\]

\item $a+c<b$. In the parameterization (\ref{dg1par-2}) after
changing variables $H=\frac{\eta}{2}+l$, and $V=\frac{\eta}{2}+m$
take the limit $\eta\to \infty$ keeping $\xi=\lambda-\eta$ fixed.
The weights will converge (up to a common factor) to:
\[
a_1:a_2:b_1:b_2:c_1:c_2\to
(e^\xi-e^{-\xi})e^{l+m}:0:e^{\xi+l-m}:e^{\xi-l+m}:1:1
\]

\end{itemize}
The two limits are related by inverting horizontal arrows. From
now on we will focus on the 5-vertex model obtained by the limit
from the 6-vertex one when $a>b+c$.

The phase diagram of the 5-vertex model is easier then the one for
the 6-vertex model but still sufficiently interesting. Perhaps the
most interesting feature is that the existence of the tricritical
point in the phase diagram.

We will use the parameter
\[
\gamma=e^{-2\lambda}
\]
Notice that $\gamma<1$.

\begin{figure}[t]
\begin{center}
\includegraphics[width=8cm, height=8cm]{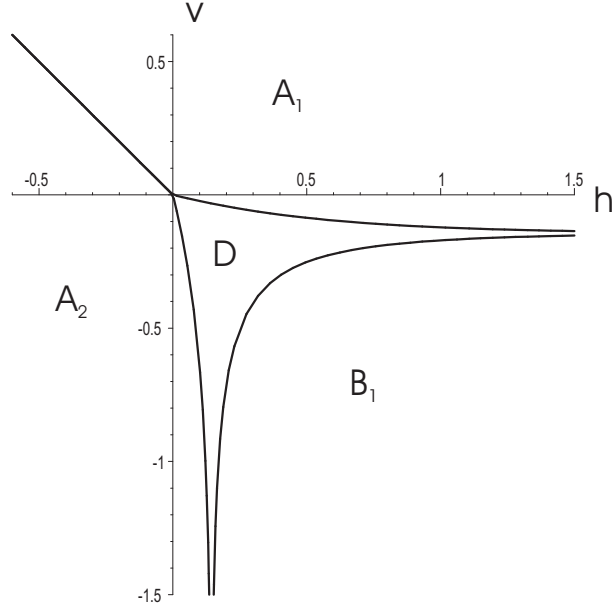}
\end{center}
\caption{The phase diagram of the $5$-vertex model with
$\gamma=1/4$ ($\beta=e^{-2h}$).} \label{amoeba_5v}
\end{figure}

The frozen regions on the phase diagram of the $5$-vertex model,
denoted on Fig. \ref{amoeba_5v} as $A_1$, $A_2$, and $B_1$, can be
described by the following inequalities:
\begin{eqnarray}
&& \mbox{$A_1$-region:}\qquad m\geq -l,\qquad l\leq 0, \nonumber
\\
&& \qquad\qquad\qquad e^{2m}\geq 1-\gamma(1-e^{-2l}), \qquad l>1;
\nonumber
\\
&& \mbox{$A_2$-region:}\qquad m\leq -l, \qquad l\leq 0,
\\
&& \qquad\qquad\qquad e^{2m}\leq 1-\frac{1}{\gamma}(1-e^{-2l}),
\qquad l>1; \nonumber
\\
&& \mbox{$B_1$-region:}\qquad
(e^{2l}-\frac{1}{1-\gamma})(e^{-2m}-\frac{1}{1-\gamma})\geq
\frac{\gamma}{(1-\gamma)^2}, \qquad e^{2l}>\frac{1}{1-\gamma};
\nonumber
\end{eqnarray}

\subsection{The asymptotic of the free energy near the
tricritical point in the 5-vertex model }

The disordered region $D$ near the tricritical point forms a
corner
$$
-\frac{1}{\gamma}\,l+O(l^2)\leq m\leq -\gamma\,l+O(l^2), \qquad
h\rightarrow 0+.
$$
The angle $\theta$ between the boundaries of the disordered region
at this point is given by
$$
\cos(\theta)=\frac{2\gamma}{1+\gamma^2}.
$$

One can argue that the finiteness of the angle $\theta$ manifests
the presence of interaction in the model. In comparison,
translationary invariant dimer models can only have cusps as such
singularities.

As it follows from results \cite{HWKK} the limit from the 6-vertex
model to the 5-vertex model commutes with the thermodynamical
limit and for the free energy of the $5$-vertex model we can use
the formula
\begin{equation}
\label{free_en_5v} f(l,m)=\lim_{\eta\rightarrow+\infty}
(F(\eta/2+l,-\eta/2+m)-F(\eta/2,-\eta/2)),
\end{equation}
where $F(H,V)$ is the free energy of the $6$-vertex model.

\begin{theorem}
Let $\gamma\leq k\leq \frac{1}{\gamma}$ and
$$
m=-kl,
$$
The asymptotics of the free energy along this ray inside the
corner near the tricritical point is given by
\begin{equation}
f(l,-kl)=c_1(k,\gamma)l +c_2(k,\gamma)l^{5/3}+O(l^{7/3}).
\end{equation}
where
\begin{equation}
c_1(k,\gamma)=\frac{1}{1-\gamma}\left(-(1+k)(1+\gamma)+4\sqrt{k\gamma}\right),
\end{equation}
and
\begin{equation}
c_2(k,\gamma)=(6\pi)^{2/3}\frac{2\gamma^{5/6}(1-\gamma)
k^{3/2}(\sqrt{k}-1/\sqrt{\gamma})^{4/3}}{5(\sqrt{k}-\sqrt{\gamma})^{4/3}}.
\end{equation}
\end{theorem}

The proof is computational. The details can be found in \cite{P}.

The scaling along any ray inside the corner near the tricritical
point in the 6-vertex model differ from this only by in
coefficients. The exponent $h^{5/3}$ is the same. The details will
be given in a separate publication.

\subsection{The limit $\Delta\to -1^-$}
If $m=1$, the region $A$ consists of one point located at the
origin. When $\eta\rightarrow 0+$ we have $\Delta\rightarrow -1-$
and $m\rightarrow 1-$. Moreover, $K'\rightarrow \pi/2$,
$K\rightarrow \frac{\pi^2}{2\eta}$, and $\frac{1}{{\rm
dn}(u|1-m)}\sim 1+\frac{1}{2}(1-m)\sin^2(u)$. Since
$\cosh^{-1}(x)\sim \pm\sqrt{2(x-1)}$, when $x\rightarrow 1+$, we
have
$$
\theta_0=\frac{|b-a|\eta}{a+b}.
$$
Since $\Xi(\varphi)$ is an odd function we obtain the following
asymptotic of $\Xi$ \cite{LW}:
$$
\Xi(\varphi)\sim 4e^{-\frac{\pi^2}{2\eta}}
\sin(\frac{\pi}{2\eta}\varphi).
$$
In this limit the antiferromagnetic region degenerates into the
origin $H=0$ and $V=0$ exponentially fast. We note that the point
$H=V=0$ is special for $|\Delta|<1$.


\subsection{The convexity} The following identity holds in the region $D$
\cite{NK}:
\begin{equation}\label{hessian}
f_{H,H}f_{V,V}-f_{H,V}^2=\left(\frac{2}{\pi g}\right)^2.
\end{equation}

Here $g=\frac{1}{2D_0^2}$. The constant $D_0$ does not vanish in
the $D$-region including its boundary. It is determined by the
solution to the integral equation for the density $\rho(z)$(see
Appendix C).

Directly from the definition of the free energy we have
\[
f_{H,H}=\lim _{N,M\to \infty}\frac{<(n(L)-n(R))^2>}{NM},
\]
where $n(l)$ and $n(r)$ are the number of arrows pointing to the
left and the number of arrows pointing to the right, respectively.

Therefore, the matrix  $\pa_i\pa_j f$ of second derivatives with
respect to $H$ and $V$  is positive definite.

As it follows from the asymptotical behavior of the free energy
near the boundary of the $D$-phase, despite the fact that the
Hessian is nonzero and finite at the boundary of the interface,
the second derivative of the free energy in the transversal
direction at a generic point of the interface develops a
singularity.

\section{The Legendre Transform of the Free Energy}
The Legendre transform of the free energy
$$
\sup_{H,V}\Bigl(xH+yV+f(H,V)\Bigr)
$$
as a function of $(x,y)$ is defined for $-1\leq x , y\leq 1$.

The variables $x$ and $y$ are known as polarizations and are
related to the slope of the Gibbs measure as $x=2h-1$ and
$y=2v-1$. We will write the Legendre transform of the free energy
as a function of $(h,v)$
\begin{equation}\label{surface_tension}
\sigma(h,v)=\sup_{H,V}\Bigl((2h-1)H+(2v-1)V+f(H,V)\Bigr).
\end{equation}
$\sigma(h,v)$ is defined on $0\leq h,v\leq 1$.

For the periodic boundary conditions the surface tension function
has the following symmetries:
$$
\sigma(x,y)=\sigma(y,x)=\sigma(-x,-y)=\sigma(-y,-x).
$$
The last two equalities follow from the fact that if all arrows
are reversed, $\sigma$ is the same, but the signs of $x$ and $y$
are changed. It follows that $\sigma_h(h,v)=\sigma_v(v,h)$ and
$\sigma_v(h,v)=\sigma_h(v,h)$.

The function $f(H,V)$ is linear in the domains that correspond to
conic and corner singularities of $\sigma$. Outside of these
domains (in the disordered domain $D$) we have
\begin{equation}\label{f-sig-comp}
\nabla\sigma\circ\nabla f={\rm id}_D, \ \nabla f\circ\nabla \sigma
={\rm id}_{\nabla f(D)}.
\end{equation}
Here we gradient of a function as a mapping $\rr^2\to \rr^2$.

When the 6-vertex model is formulated in terms of the height
function, the Legendre transform of the free energy can be
regarded as a surface tension. The surface in this terminology is
the graph of of the height function.

\subsection{}
Now let us  describe some analytical properties of the function
$\sigma(h,v)$ is obtained as the Legendre transform of the free
energy. The Legendre transform maps the regions where the free
energy is linear with the slope $(\pm 1, \pm 1)$ to the corners of
the unit square ${\mathcal D}=\{(h,v)|\quad 0\leq h\leq 1, 0\leq
v\leq 1 \}$. For example, the region $A_1$ is mapped to the corner
$h=1$ and $v=1$ and the region $B_1$ is mapped to the corner $h=1$
and $v=0$. The Legendre transform maps the tentacles of the
disordered region to the regions adjacent to the boundary of the
unit square. For example, the tentacle between $A_1$ and $B_1$
frozen regions is mapped into a neighborhood of $h=1$ boundary of
$\mathcal D$, i.e. $h\to 1$ and $0<v<1$.

Applying the Legendre transform to asymptotics of the free energy
in the tentacle between $A_1$ and $B_1$ frozen regions we get
$$
H(h,v)=-\frac{1}{2}\ln\left(\frac{\pi
ab}{c^2}\frac{1-h}{\sin{\pi(1-v)}}\right),\qquad
V(h,v)=\frac{1}{2}\ln(b/a)+\frac{\pi}{2}(1-h)\cot(\pi(1-v)),
$$
and
\begin{equation}\label{sur_ten_boundary}
\sigma(h,v)=(1-h)\ln\left(\frac{\pi
ab}{c^2}\frac{1-h}{\sin(\pi(1-v))}\right) -(1-h)+v\ln(b/a)-\ln(b),
\end{equation}
Here $h\to 1-$ and $0<v<1$. From (\ref{sur_ten_boundary}) we see
that $\sigma(1,v)=v\ln(b/a)-\ln(b)$, i.e. $\sigma$ is linear on
the boundary $h=1$ of $\mathcal D$. Therefore, its asymptotics
near the boundary $h=1$ is given by
$$
\sigma(h,v)=v\ln(b/a)-\ln(b)+(1-h)\ln(1-h)+O(1-h),
$$
as $h\to 1-$ and $0<v<1$. We note that this expansion is valid
when $(1-h)/\sin(\pi(1-v))\ll 1$.

Similarly, considering other tentacles of the region $D$, we
conclude that the surface tension function is linear on the
boundary of $\mathcal D$.

\subsection{}
Next let us find the asymptotics of $\sigma$ at the corners of
$\mathcal D$ in the case when all points of the interfaces between
frozen and disordered regions are regular, i.e. when $\Delta<1$.
We use the asymptotics of the free energy near the interface
between $A_1$ and $D$ regions
(\ref{asympt_free_energy_interface}).

First let us fix the point $(H_0,V_0)$ on the interface and the
scaling factor $r$ in (\ref{asympt_free_energy_interface}). Then
from the Legendre transform we get
$$
1-h=-\frac{3}{4}\,r\,\frac{\kappa(\theta s+t^2)^{1/2}}{(\partial_H
g)^2+4} (\theta\partial_H g+4rt)
$$
and
$$
1-v=-\frac{3}{2}\,r\,\frac{\kappa(\theta s+t^2)^{1/2}}{(\partial_H
g)^2+4} (-\theta+r\partial_H g t).
$$
It follows that
$$
\frac{1-h}{1-v}=\frac{\theta\partial_H g
+4rt}{2(-\theta+rt\partial_H g)}.
$$
In the vicinity of the boundary $r\to 0$ and, hence,
\begin{equation}\label{slope_corner}
\frac{1-h}{1-v}=-\frac{\partial_H g }{2}= \frac{1-b/a\,\,
e^{-2V_0}}{1-b/a\,\, e^{-2H_0}}
\end{equation}
as $h,v\to 1$. Thus, under the Legendre transform, the slope of
the line which approaches the corner $h=v=1$ depends on the
boundary point on the interface between the frozen and disordered
regions.

It follows that the first terms of the asymptotics of $\sigma$ at
the corner $h=v=1$ are given by
$$
\sigma(h,v)=-\ln a-2(1-h)H_0(h,v)-2(1-v)V_0(h,v),
$$
where $H_0(h,v)$ and $V_0(h,v)$ can be found from
(\ref{slope_corner}) and $g(H_0,V_0)=0$.

When $|\Delta|\leq 1$ the function $\sigma$ is strictly convex and
smooth for all $0<h,v<1$. It develops conical singularities near
the boundary.

When $\Delta <-1$, in addition to the singularities on the
boundary, $\sigma$ has a conical singularity at the point
$(1/2,1/2)$. It corresponds to the ``central flat part'' of the
free energy $f$, see Fig. \ref{antiferroelectric_diagram}.

When $\Delta >1$ the function $\sigma$ has corner singularities
along the boundary as in the other cases. In addition to this, it
has a corner singularity along the diagonal $v=h$ if $a>b$ and
$v=1-h$ if $a<b$. We refer the reader to \cite{BS} for further
details on singularities of $\sigma$ in the case when $\Delta>1$.

\section{The Thermodynamic Limit and the Variational Principle for
Fixed Boundary Conditions} \label{section_var}

\subsection{The Variational Principle}
\subsubsection{}Let $\sigma$ be the surface tension function of the $6$-vertex model with the
periodic boundary conditions defined in (\ref{surface_tension}).
We consider the functional
\begin{equation}\label{var-pr}
I[\varphi]=\int_D\sigma(\nabla \varphi)d^2x+\lambda\int_D \varphi
d^2x,
\end{equation}

Let $h(x,y)$ be a minimizer of this functional on the space
$L(D,\varphi_0)$ of functions nondecreasing in $x$ and $y$
directions and satisfying the condition
$$
\varphi(x,y)-\varphi(x',y')\leq x-x'+y-y'
$$
and the boundary condition
$$
\varphi|_{\partial D}=\varphi_0.
$$
Notice that height functions are Lipschitz with
$|\varphi(x,y)-\varphi(x',y')| \leq 2 max(| x-x'|,|y-y'|)$.

\begin{proposition}The functional $I[\varphi]$ has unique minimizer.
\end{proposition}

Indeed, since $\sigma$ is convex, the minimizer is unique when it
exists. The existence of the minimizer follows from compactness of
the space $L(D,\varphi_0)$ in the sup norm. The arguments are
completely parallel to those in \cite{CKP}.

\subsubsection{}If the vector $\nabla h(x,y)$ is not a singular point of $\sigma$,
the minimizer $h$ satisfies the Euler-Lagrange equation in a
neighborhood of $(x,y)$
\begin{equation}\label{E-L_eq}
{\rm div}(\nabla\sigma\circ\nabla h)=\lambda.
\end{equation}
We can also rewrite this equation in the form
\begin{equation}\label{fun_g}
\nabla\sigma(\nabla
h(x,y))=\frac{\lambda}{2}\,(x,y)+(-g_y(x,y),g_x(x,y)),
\end{equation}
where $g$ is an unknown function such that
$g_{xy}(x,y)=g_{yx}(x,y)$. It is determined by the boundary
conditions for $h$.

Applying (\ref{f-sig-comp}), we see that
\begin{equation}
\nabla h(x,y)=\nabla f\Bigl(\frac{\lambda}{2}\,x-g_y(x,y),
\frac{\lambda}{2}\,y+g_x(x,y)\Bigr). \label{height}
\end{equation}

From the definition of the slope, see (\ref{slope}) we have $0\leq
f_H \leq 1$ and $0\leq f_V\leq 1$. Thus, if the minimizer $h$ is
differentiable at $(x,y)$, it satisfies the constrains $0\leq h_x
\leq 1$ and $0\leq h_y\leq 1$.

\subsection{Large deviations}
The following statement  is a minor variation of the theorem 4.3
from \cite{CKP}.

\begin{theorem} Let $N\to \infty$, $\lambda$ be finite and
$q=\exp(\frac{\lambda}{N})$ then the sequence of random normalized
height functions $h_N$ converges  in probability  to the minimizer
of (\ref{var-pr}). The rate of convergence is exponential of
$N^2$.
\end{theorem}

The minimizer of the variational problem (\ref{var-pr}) is called
the {\it limit shape} of the height function.

This theorem is the manifestation of the general philosophy of the
large deviations principle.  The probability of a having state
with the height function $h$ has the has the following asymptotic
$N\to \infty$ :
$$
Prob(h)\sim\exp\Bigl(\lambda N^2\int_D h d^2x
+N^2\int_D\sigma(\nabla h)d^2x\Bigr).
$$
Here $\sigma$ is the surface tension function for the periodic
boundary conditions. Clearly this probability has maximum at the
limit shape. States with the height function, which
macroscopically differ from the limit shape, should be expected to
be exponentially improbable. The theorem states that this is
exactly what is taking place.

\section{The limit $|\lambda|\rightarrow\infty$ }

\subsection{Minimal and maximal height functions}
The space of normalized height functions on $L_{N,M}$ has the
partial order described in section 2. Denote the minimal and
maximal normalized height functiions with respect to this partial
order  $h^{min}$ and $h^{max}$, respectively.

For two functions $h_1$, $h_2$ on $D$ define the distance
\begin{equation}\label{distance}
dist(h_1, h_2)=sup_{x\in D}|h_1(x)-h_2(x)|
\end{equation}
Similarly define the the distance between two functions on the
boundary of $D$.

Let $h^{(N)}_0$ be a sequence of functions on the boundary of $D$
converging to $\varphi_0$, and such that $h^{(N)}_0$ is a boundary
of a normalized  height function of $L_{N,M}$. Denote
$h_{min}^{(N)}$ and $h_{max}^{(N)}$ the minimum and maximum height
functions from $L_{N,M}(h^{(N)}_0)$. The following is clear:

\begin{proposition}Let $\varphi_{min}$ and  $\varphi_{max}$
be minimal and maximal height functions with the boundary
condition $\varphi_0$. Then,
$\varphi_{min}=\lim_{N\rightarrow\infty}h_N^{min}$ and
$\varphi_{max}=\lim_{N\rightarrow\infty}h_N^{max}$ with respect to
the distance (\ref{distance}).
\end{proposition}

The functions $\varphi_{min}$ and  $\varphi_{max}$ minimize the
functionals
$$
\pm \int_D \varphi d^2x
$$
in the sapec $H(\varphi_0)$.

Let us assume that the boundary conditions are critical, that is
$\varphi_0$ is piece-wise linear, non-decreasing in $x$ and $y$
direction with the slope $0$ or $1$. In this case the minimum and
maximum height functions are piecewise linear functions, such that
each linear part has the slope  either $0$ or $\pm 1$ along
coordinate axes.

We will say a point $(x,y)$ is \emph{regular} if at this point the
function $\varphi_{min}$ is differentiable. For critical boundary
conditions regular points form regions with piece-wise linear
boundary where the function $\varphi_{min}$ has a constant slope.
We will call them \emph{linear domains}.

Points where the gradient of $\varphi_{min}$ is discontinuous will
be called \emph{singular} points. For critical boundary conditions
singular points are the points where several linear domains meet.
The {\emph valency} of a singular point is the number of linear
domains which meet at this point. For generic critical boundary
conditions the valency of each critical point is at most three.

\begin{figure}[t]
\begin{center}
\includegraphics[width=8cm, height=10cm]{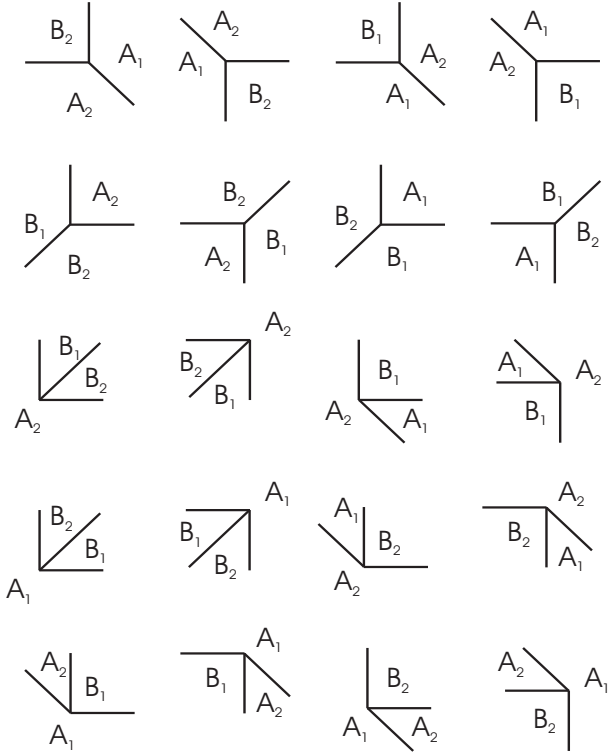}
\end{center}
\caption{Possible phases at a tricritical point of the $6$-vertex
model.} \label{triple_pts}
\end{figure}

The list of all possible phases at a tricritical point of the
$6$-vertex model with generic critical boundary conditions is
given on Fig. \ref{triple_pts}.

\subsection{The asymptotic of the minimizer when $|\lambda|\rightarrow\infty$}
Here we study the asymptotic of minimizer of $I_\lambda[\varphi]$
as $\lambda\to \pm \infty$. It is more convenient to divide
$I_\lambda$ by $|\lambda|$, so we are looking for the asymptotics
of the minimizer $h_\lambda$ of
$$
I^{\pm}_\lambda[\varphi]=\frac{1}{|\lambda|}\int_D
\sigma(\nabla\varphi)d^2x \pm \int_D \varphi d^2x.
$$

Let us focus on the limit $\lambda\rightarrow +\infty$. The limit
$\lambda\rightarrow -\infty$ can be treated similarly.

When $\lambda\to +\infty$, the minimizer $h_\lambda$ approaches
the minimal height function $\varphi_{min}$ described above. Let
us look for the asymptotical formula for the minimizer in a small
neighborhood of a point $(x_0,y_0)$ of the form
$$
h_\lambda(x,y)=\varphi_{min}(x,y)
+\frac{1}{\lambda}\,\,H_\lambda\bigl(\lambda(x-x_0),
\lambda(y-y_0)\bigr),
$$
where $H_\lambda=H+o(1)$ as $\lambda\to +\infty$.

Let $z=(z_1,z_2)=(r,s)=(\lambda(x-x_0),\lambda(y-y_0))$ and
$x=(x_1,x_2)=(x,y)$. Because $\varphi_{min}$ is linear, its second
derivatives vanish and we can rewrite the Euler-Lagrange equation
(\ref{E-L_eq}) as
\begin{equation}\label{EL-H}
\frac{\pa^2 H_\lambda}{\pa z_i\pa z_j} \frac{\pa^2\sigma}{\pa
u_i\pa u_j}\bigl(\nabla_{x}\,\,\varphi_{min}(x,y)+
\nabla_{z}\,\,H_\lambda(z)\bigr)=1.
\end{equation}

Notice that for critical boundary conditions the function $\nabla
\varphi_{min}$ is piece-wise constant.

Now assume that $(x_0,y_0)$ is a singular point, i.e. a point
where two or more linear domains of $\varphi_{min}$ meet. Recall
that for generic critical boundary conditions only two or three
linear domains can meet at a point. Taking the limit $\lambda\to
\infty$ in (\ref{EL-H}) we obtain.

\begin{proposition}
Let $h_\lambda$ be the minimizer of $I^+_{\lambda}$ and
$(x_0,y_0)$ be a singular point, then for each $(s,t)\in \rr^2$
there exits
\begin{equation}\label{asympt_fun}
H(r,s)=\lim_{\lambda\to\infty}\lambda
\left(h_\lambda(x_0+\frac{r}{\lambda}, y_0+\frac{s}{\lambda})-
\varphi^{min}(x_0,y_0)\right)
\end{equation}
This function is the solution to
\begin{equation}\label{H-eqn}
{\rm div}(\nabla\sigma\circ(\nabla H))=1,
\end{equation}
with the boundary conditions
\begin{equation}\label{as-cond}
H(r,s)\to k_ir+l_is
\end{equation}
as $(r,s)\to \infty$ along a ray in the $i$-th linear domain
adjacent to $(x_0,y_0)$. Here $(k_i,l_i)=\nabla \varphi_{min}$ in
the $i$-th linear domain and since the boundary conditions are
critical, $k_i,l_i=0, 1$.
\end{proposition}

\begin{conjecture} The function $H$ is once differentiable. There is a smooth
curve separating $\rr^2$ into regions where $H$ is linear and
regions where $H$ is smooth with positive definite matrix of
second derivatives.
\end{conjecture}

\begin{remark}
For non-generic boundary conditions in the thermodynamical limit
more then three linear domains can met at one point. In this case
one should expect that the conjecture still holds with more then
three linear domains meeting at a singular point.
\end{remark}

\subsection{The asymptotic near double degenerate singular points }
\subsubsection{}A height function defines the surface $z=h(x,y)$ in
$\rr^3$. Regions where $h$ is linear are planes. Here we will
describe the solution to the equation (\ref{H-eqn}) with the
boundary conditions (\ref{as-cond}) in the case when there are
only two asymptotic planes meeting at$(x_0,y_0)$  .

Let $(k_1,k_2)$ and $(l_1,l_2)$ be the directions of the steepest
assent of these planes. The numbers $k_i$ and $l_i$ are either $0$
or $ 1$ since we assume critical boundary conditions and therefore
$\varphi_{min}$ is piece-wise linear with slope $(0,0),
(0,1),(1,0)$ or $(1,1)$.

The function $H(r,s)$, defined in (\ref{asympt_fun}), has the
asymptotic conditions (\ref{as-cond})determined by these planes.
The function $H(r,s)$ is also invariant with respect to
translations in $m=(k-l)^{\perp}$-direction.

Thus, we are looking for a function $\kappa$ such that
\[
H(r,s)=\kappa\left((k_1-l_1)r+(k_2-l_2)s\right),
\]
which satisfies the differential equation (\ref{H-eqn}) with the
asymptotic conditions (\ref{as-cond}).

Let us introduce
\[
S(t)=\sum_{i,j=1,2}(k-l)_i(k-l)_j\frac{\partial^2\sigma}{\partial
u_i\partial u_j}(u)\Bigl|_{u=(k-l)t}.
\]
Then the differential equation for $H$ becomes the first order ODE
for the function $\kappa'$
\begin{equation}\label{eqn-for-kappa}
\kappa''(t)S(\kappa'(t))=1.
\end{equation}
Integrating it, we obtain the equation defining the function
$g(t)=\kappa'(t)$ implicitly
\begin{equation}\label{implic_eq_g}
\sum_{i=1,2} (k_i-l_i)\frac{\partial\sigma}{\partial
u_i}(g(t)(k-l))=t+C
\end{equation}
with some constant $C$.

\subsubsection{}
\begin{figure}[b]

\begin{center}
\includegraphics[width=8cm, height=4cm]{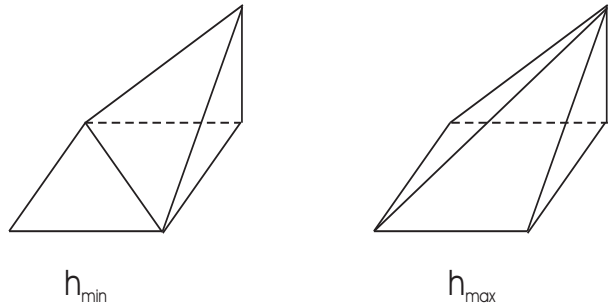}
\end{center}
\caption{The minimum and maximum height functions for the DW
boundary conditions.} \label{DW_min_max}
\end{figure}

For the domain wall boundary conditions minimal and maximal height
functions are shown on Fig. \ref{DW_min_max}. In this case the
boundary between two linear domains is a line in the direction
$(1,-1)$ for the minimal height function and is the line in the
direction $(1,1)$ for the maximal height function.

Thus, for these boundary conditions the function $H$ describing
the asymptotic of the minimizer is constant in the direction
$(1,-1)$ when $\lambda\to +\infty$ and it is constant in the
$(1,1)$-direction if $\lambda\to -\infty$.

Let $H(r,s)=\kappa(s+r)$ be the solution to (\ref{eqn-for-kappa})
which is invariant with respect to translations in the $(1,-1)$
direction. The symmetries of the Legendre transform $\sigma$ of
the free energy  imply that $\sigma_h(h,v)=\sigma_v(v,h)$. Using
the equation (\ref{implic_eq_g}) and this symmetry of $\sigma$, we
obtain
$$
\frac{\pa \sigma}{\pa u_1}(g(t),g(t))=t/2+C/2.
$$
Taking into account (\ref{f-sig-comp}) we obtain
$$
g(t)=\partial_H f(t/2+C/2,t/2+C/2) =\partial_V f(t/2+C/2,t/2+C/2)
$$
and, hence,
$$
\kappa(t)=f(t/2+C/2,t/2+C/2).
$$

In the case of two asymptotic planes we have an additional
symmetry of $H(r,s)$ with respect to the intersection line of
these planes, i.e. $\kappa(t)=\kappa(-t)$. The free energy $f$
also has the symmetry $f(x,y)=f(-y,-x)$. Therefore $C=0$ and we
proved the following statement.

\begin{theorem} For DW boundary conditions
the minimizer $h_\lambda$ when $\lambda\to+\infty$ has the
following asymptotic when $x=x_0+r/\lambda$, $y=-x_0+s/\lambda$:
\begin{equation}
h_\lambda(x,y)=\varphi_{min}+\frac{1}{\lambda}\,\,
f\left(\frac{r+s}{2}\,,\frac{r+s}{2}\right)+o\Bigl(\frac{1}{\lambda^2}\Bigr).
\end{equation}

If $x=x_0+r/\lambda$ and $y=x_0+s/\lambda$ and $\lambda\to
-\infty$ the asymptotic of the minimizer $h_\lambda$ is given by
\begin{equation}
h_\lambda(x,y)=\varphi_{min}+\frac{1}{\lambda}\,\,
f\left(\frac{r-s}{2}\,,-\frac{r-s}{2}\right)
+o\Bigl(\frac{1}{\lambda^2}\Bigr).
\end{equation}
\end{theorem}

\subsubsection{} When $c>a+b$ (which is equivalent to $\Delta<0$) the height function
$h_\lambda$ develops extra linear domains known as facets. These
are the regions of the antiferroelectric phase discussed in
section \ref{AF-region}. In dimer models this phenomenon is
studied in \cite{KOS}.

The facets also appear in the function $H$ describing the
asymptotic of the minimizer as $\lambda\to \infty$.

For the function $H$ describing the asymptotic near a double
singular point $(x_0,y_0)$ where two linear domains meet along the
diagonal $x+y=0$ the facet is a strip
$$
|s+t|\leq R,
$$
in coordinates $x=x_0+\frac{s}{\lambda},
y=-x_0+\frac{t}{\lambda}$. Its width $R$ is given by the formula
$$
R=\sqrt{2}\,\,\Bigl|\Xi(\frac{\eta+\theta_0}{2})\Bigr|.
$$

In the limit $\Delta\rightarrow -1-$ or $\eta\rightarrow 0+$ the
asymptotic of $\Xi$ gives the following asymptotical value of $R$:
\begin{equation}
R=4 \sqrt{2} e^{-\frac{\pi^2}{2\eta}}
\sin\Bigl(\frac{\pi\max(a,b)}{2(a+b)}\Bigr)(1+O(\eta)).
\end{equation}

The free energy $f$ is linear in the antiferromagnetic region. It
is growing with exponent $3/2$ in the normal direction to the
boundary outside of the antiferromagnetic region. This agrees with
the Pokrovsky-Talapov law which states that $f$ should be growing
with exponent $3/2$ in the normal direction to the boundary of the
facet \cite{PT}. In particular, $h$ grows with the exponent $3/2$
in the $(1,1)$-direction near the boundary of the $c$-droplet far
enough from the boundary of the square. For $s\to R+0$ we have
$$
h_\lambda(x_0+\frac{s}{\lambda},-x_0)=h_0
+\frac{\kappa}{\lambda}(s-R)^{3/2},
$$
where $h_0$ is the value of the height function at the boundary of
the facet and $\kappa$ is a constant which can be computed
explicitly.

\section{Conclusion}

\subsection{Correlation functions in the bulk} As we have seen in the previous section at the
macroscopical distances in the thermodynamical limit the height
function is deterministic and is the minimizer for the variational
problem (\ref{var-pr}).

But the height function at smaller distances remain random. Their
fluctuations are described by the asymptotical behavior of
correlation functions in the thermodynamical limit. These
asymtotics have been studied extensively in dimer models which
describe the $\Delta=0$ case of the 6-vertex model.

The thermodynamical limit of correlations functions in the bulk
describes translationary invariant Gibbs measure with given
polarization. These asymptotics of correlation functions have been
studied a lot using various methods which are essentially based on
representation theory of affine quantum groups.

The exact computation of the asymptotic of correlation functions
for local observable separated by large distances on the lattice
remain one of the main problems. On the other hand this is one of
the most important physically relevant information about
correlation functions . Some information about this asymptotic of
correlation functions can be obtained using the arguments of
finite-size scaling and the assumption of conformal invariance of
the leading terms of the asymptotic.

Let us consider the 6-vertex configurations of paths which may end
at some edges. The weight of such configurations is given the same
product of weights as before. Define local observables $\tau_e$
and $\tau_e^*$ as
\[
\tau_e(S)=\left\{ \begin{array}{cc} 1,
&\mbox{if a path going up starts at} \ e ; \\
0, &  \mbox{otherwise}  \end{array} \right .
\]
\[
\tau^*_e(S)=\left\{ \begin{array}{cc} 1,
&\mbox{if a path going down starts at} \ e ; \\
0, &  \mbox{otherwise}  \end{array} \right .
\]
The value of a product of such observable when each of the factors
correspond to a different edge is the product of values of
observables.

The following formulae were obtained in \cite{BIR} for $H=0$ and
when all edges are vertical on the same row using the finite-size
scaling and the assumption of conformal invariance:
\begin{equation}\label{as-1}
<\sigma_{e_1}\sigma_{e_2}>\simeq <\sigma_e>^2+\frac{A}{d^2}+
\frac{Bcos(2k_F d)}{d^\alpha}+\dots
\end{equation}
\begin{equation}\label{as-2}
<\tau^*_{e_1}\tau_{e_2}>\simeq
\frac{C}{d^{\frac{1}{\alpha}}}+\dots
\end{equation}
Here $j$ is the distance between $e_1$ and $e_2$,
$\alpha=2\pi^2\rho(\xi)^2$. The terms denoted by $\dots$ are with
higher powers of $d^{-2}$ and $d^{-\alpha}$ where $d$ is the
distance between $e_1$ and $e_2$.  These higher order terms may
also be oscillating. Because $H=0$ the integration contour $C$ in
the integral equation fro $\rho(z)$ is a segment of the real line
in the additive parameterization. The constant $k_F$ is the
Fermi-momentum and it is also can be expressed in term of
$\rho(z)$. It is also equal to the vertical electrical
polarization.

The finite size computations were extended to the case $V\neq 0$
in \cite{NK}. They argued  that the complete spectrum of effective
$c=1$ conformal field theory is given by
\[
\Delta_\pm =\frac{1}{4} (\frac{m^2}{g}+n^2g\pm 2nm), \ \ n,m\in
\zz
\]
where $g$ is defined by the Hessian of the free energy  and is
related to $\alpha$ as $\alpha=\frac{1}{2g}$.

Combining these results we conjecture that for generic electric
fields the asymptotical behavior of correlation functions is still
given by formulae (\ref{as-1}) and (\ref{as-2}) with
$\alpha=\frac{1}{2g}$

One can show easily that  when $\Delta=0$ we have  $g=1/4$ and the
asymptotics (\ref{as-1}) (\ref{as-2}  agrees with the results from
\cite{K} on dimer models.

\subsection{Open problems}
Here we will list some open question about the 6-vertex model and
other related models.

\begin{itemize}

\item Find the classification of generic singularities of limit
shapes. For $\Delta>1$ one should expect the presence of corners
as generic singularities of limit shapes.

\item Find the scale at which fluctuations near singularities of
limits shapes are described by some random process and describe
such processes. For example in dimer models such fluctuations near
the boundary of the limit shape are described by Airy process, and
near a generic cusp are described by the Pearcey process.

One can argue that the same processes should describe correlation
functions near similar singularities of the limit shape in the
6-vertex model but this is still a conjecture.

The scaling near the corner singularity seems particularly
interesting problem since such singularities do not appear in
dimer models.

\item Understand the role of integrability of the 6-vertex model
in the formation of the limit shape. Here by integrability we mean
that the model can be solved by the Bethe ansatz, and that the
weights satisfy the Yang-Baxter equation (and therefore
transfer-matrices form a commuting family).

\item The 6-vertex model is closely related to the representation
theory of quantized universal enveloping algebra of
$\widehat{sl_2}$. It would be extremely interesting to see which
aspect of the representation theory of this algebra naturally
appear in the limit shape phenomenon and in the scaling of
correlation functions near singularities.

\item The 6-vertex model has natural generalizations related to
other simple Lie algebras. Configurations in these models can be
described in terms of $r$ height functions where $r$ is the rank
of the Lie algebra. The limit shape in this case is a surface in
$\rr^{r+2}$. It would be extremely interesting to investigate such
systems trying to answer all questions mentioned above.

\end{itemize}

\appendix

\section{The Partition Function of the Inhomogeneous $6$-Vertex Model}
\label{a_part_fun_inhom} Here we compute the partition function of
the inhomogeneous $6$-vertex model with the periodic boundary
conditions applying the Bethe Ansatz method. We briefly describe
the results and refer the reader to \cite{KBI} for the details
about the method.

\begin{figure}[b]
\begin{center}
\includegraphics{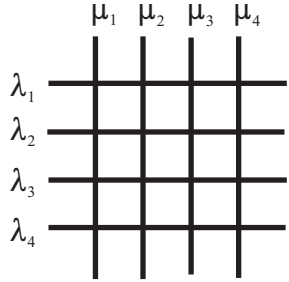}
\end{center}
\caption{The $4$-by-$4$ square grid with corresponding spectral
parameters.} \label{spec_par}
\end{figure}

Bethe Ansatz method works for the inhomogeneous 6-vertex model
with the follwing inhomogeneities. One should parameters
$\lambda_1$, $\lambda_2$,.., $\lambda_M$ to each row of the grid
$L_{N,M}$ and parameters $\mu_1$, $\mu_2$,.., $\mu_N$ to each
column of the grid as shown on Fig. \ref{spec_par}. Thus, a pair
of spectral parameters $(\lambda_i,\mu_j)$ is associated to the
vertex $(i,j)$ of the grid. Boltzmann weights assigned to this
point are:
\[
a_1(i,j)=a(\lambda_i-\mu_j)e^{H+V}, \
a_1(i,j)=a(\lambda_i-\mu_j)e^{H+V}
\]
\[
b_1(i,j)=b(\lambda_i-\mu_j)e^{H-V}, \
b_1(i,j)=b(\lambda_i-\mu_j)e^{-H+V}
\]
\[
c_1(i,j)=c(\lambda_i-\mu_j), \ c_1(i,j)=c(\lambda_i-\mu_j)
\]
where functions $a(\lambda), b(\lambda), c(\lambda)$ describe the
parameterization of Boltzmann weights of the homogeneous model.

The partition function of the model with periodical boundary
conditions can be written as
$$
Z_N={\rm Tr}\prod_{k=1}^N T(\lambda_k),
$$
where $T$ is the $2^N\times 2^N$ row-to-row transfer matrix of the
6-vertex model \cite{Bax}. The raw-to-raw transfer-matrix is the
trace of the "quantum monodromy matrix":
$$
T(\lambda_k)={\rm tr_0}\,\tau(\lambda_k).
$$
Here $\tau$ is the product
$$
\tau(\lambda_k)=R_{0,1}(\lambda_k-\mu_1)\dots
R_{0,N}(\lambda_k-\mu_N) ,
$$
which acts in $\cc^2\otimes {\cc^2}^{\otimes N}$. The matrix
$R_{0,j}(\lambda_k-\mu_j)$ acts as the matrix
$$
\begin{pmatrix}
a_1(\lambda_k-\mu_j)&0&0&0\\
0&b_2(\lambda_k-\mu_j)&c_1(\lambda_k-\mu_j)&0\\
0&c_2(\lambda_k-\mu_j)&b_1(\lambda_k-\mu_j)&0\\
0&0&0&a_2(\lambda_k-\mu_j)
\end{pmatrix}.
$$
in the basis $e_1\otimes e_1,e_1\otimes e_2,e_2\otimes
e_1,e_2\otimes e_2,$ of the product of the 0-th and j-th factors
in the tensor product. It acts trivially in other factors.

Let us denote the number of arrows pointing down in a row of the
grid by $n$. Denote by $(\cc^{\otimes N})_n$ the corresponding
subspace in the space of all possible states on the raw of length
$N$ of vertical edges. The action of the transfer-matrix preserve
these subspaces.

The Bethe Ansatz method gives the following result for the
eigenvalues of $T$:
\begin{equation}
\Lambda(\lambda, u_1,.., u_n, H, V)= \Lambda_L(\lambda, u_1,..,
u_n, H, V)+ \Lambda_R(\lambda, u_1,.., u_n, H, V),
\end{equation}
where
\begin{eqnarray}
&& \Lambda_R(\lambda, u_1,.., u_n, H, V)=
e^{NH+(N-2n)V}\prod_{l=1}^n
\frac{a(u_l-\lambda)}{b(u_l-\lambda)}\prod_{j=1}^N
a(\lambda-\mu_j), \nonumber
\\
&& \Lambda_L(\lambda, u_1,.., u_n, H, V)=
e^{-NH+(N-2n)V}\prod_{l=1}^n \frac{a(\lambda- u_l)}{b(\lambda-
u_l)}\prod_{j=1}^N b(\lambda-\mu_j),
\end{eqnarray}
and $u_i$ satisfy the Bethe equations
\begin{equation}
e^{2NH}\prod_{k=1,k\neq l}^n
\frac{a(u_k-u_l)b(u_l-u_k)}{a(u_l-u_k)b(u_k-u_l)}
=\prod_{j=1}^N\frac{b(u_l-\mu_j)}{a(u_l-\mu_j)}. \label{Bethe}
\end{equation}

It follows that the partition function of the model is given by
$$
Z_{N,M}=\sum_{a}\prod_{k=1}^N
\Lambda_a(\lambda_k,u_1,u_2,..,u_n,H,V).
$$
where $\Lambda_a$ are the eigenvalues of the row-to-row transfer
matrix $T$ and the sum is taken over all eigenvalues with their
multiplicities.

Thus, in the homogeneous case when all $\lambda_k$ are the same
the asymptotic of the partition function as $M\to \infty$ is
\[
Z_{N,M}=d\Lambda_{max}(N)^M(1+o(1))
\]
where $d$ is the multiplicity of the largest eigenvalue
$\Lambda_{max}(N)$ of the row-to-row transfer-matrix on $N$ sites
.

Therefore, if we can compute the asymptotic of the largest
eigenvalue of the transfer-matrix as $N\to \infty$ we can find the
asymptotic of the partition function in the thermodynamical limit.

Strictly speaking, this logic gives the asymptotic of the
partition function in the limit $M\gg N \gg 1$. Under mild
assumptions one can argue that the leading term of the asymptotic
of free energy is uniform. However, in some cases a resonant
phenomenon may occur which will make the leading term of the
asymptotic of $Z_{N,M}$ depend on the ratio $N/M$ \cite{KW}.

\section{The Free Energy of the Homogeneous $6$-Vertex Model}
\label{a_free_energy_hom} In the homogeneous $6$-vertex model the
Boltzmann weights are the same for all vertices on the grid $L_N$.
When $|\Delta<1$ the functions $a, b, c$ are:
$$
a=\rho\sin(u+\eta),\quad b=\rho\sin(u),\quad c=\rho\sin(\eta).
$$
Other values of $\Delta$ can b obtained by analytical
continuation.

The formula for eigenvalues of the row-to-row transfer matrix
given in the previous section becomes
$$
\Lambda^{(n)}(N)=a^Ne^{NH+(N-2n)V}
\prod_{j=1}^n\frac{\sin(u_j-u+\eta)}{\sin(u_j-u)}+
b^Ne^{-NH+(N-2n)V}\prod_{j=1}^n\frac{\sin(u_j-u-\eta)}{\sin(u_j-u)},
$$
where $(u_1,..,u_n)$ are solutions of the Bethe ansatz equations
$$
\Biggl[\frac{\sin(u_j)}{\sin(u_j+\eta)}\Biggr]^N=
e^{2NH}\mathop{\prod_{k=1}^n}_{k\neq j}
\frac{\sin(u_k-u_j+\eta)}{\sin(u_k-u_j-\eta)}, \quad j=1,..,n.
$$

Introducing new variables
$$
z_j=\frac{\sin(u_j)}{\sin(u_j+\eta)}, \quad j=1,.., n, \qquad
w=\frac{\sin(u)}{\sin(u+\eta)}
$$
we get
$$
\Lambda^{(n)}(N)=a^Ne^{NH+(N-2n)V}\prod_{j=1}^n\frac{1-2\Delta
w+wz_j}{z_j-w}+ b^Ne^{-NH+(N-2n)V}\prod_{j=1}^n\frac{1-2\Delta
z_j+wz_j}{w-z_j}
$$
where $z_i$ satisfy the Bethe equations
\begin{equation}\label{Bethe_eq_z}
z_j^N=(-1)^{n-1}e^{2NH}\prod_{k=1}^n \frac{1-2\Delta
z_j+z_jz_k}{1-2\Delta z_k+z_jz_k}.
\end{equation}

We want to find the asymptotic of the largest eigenvalue of the
transfer-matrix when $N\to \infty$. We shall do it in two steps.
First we will find the asymptotic of the largest eigenvalue among
$\Lambda^{(n)}(N)$ in the limit
\begin{equation}
N\longrightarrow \infty,\qquad n\longrightarrow\infty, \qquad
\frac{n}{N} =\alpha={\rm const},
\end{equation}
After this  we will find for which $0\leq\alpha\leq 1$ this
eigenvalue is the largest.

\section{The $6$-Vertex Model Bethe Equations}\label{Comp}
\label{a_bethe_eqs} Let us denote
\begin{equation}
e^{i\theta(z,w)}=\frac{1-2\Delta z+zw}{1-2\Delta w+zw}.
\end{equation}
Then, taking the logarithm of both sides of each Bethe equation
(\ref{Bethe_eq_z}) and dividing by $N$, we get
\begin{equation}\label{b-eqn}
\ln(z_j)=\frac{n-1}{N}i\pi+2H+\frac{i}{N}\sum_{k=1}^n\theta(z_j,z_k)
+2\pi i k_j/N, \qquad j=1,..,n,
\end{equation}
where $k_j$ are integers.

All theoretical and numerical results on the $6$-vertex model
suggest that the following is true.
\begin{conjecture}\label{main_conj}
In the limit $N\to\infty$ with $\alpha=\frac{n}{N}$ is fixed the
roots of the Bethe equations corresponding to the maximal
eigenvalue of the transfer matrix in the subspace $(\cc^{\otimes
N})_n$ concentrate on a smooth curve $C$ in the complex $z$-plane.
The curve $C$ is symmetric with respect to the complex conjugation
$z\rightarrow\bar{z}$.
\end{conjecture}

\begin{conjecture}
The maximal eigenvalue of the transfer-matrix in the subspace
$(\cc^{\otimes N})_n$ corresponds to $k_j=\frac{j}{N}$.
\end{conjecture}

These conjectures can be proven in the free-fermion case
$\Delta=0$.

The conjecture (\ref{main_conj}) implies that as $N\to \infty$ one
should expect finite limit density of $z_j$ on the contour $C$. If
$z_j\to z\in C$ the limit $\rho(z)=\lim_{N\to\infty}
\frac{i}{N(z_{j+1}-z_j)}$ is the value of the limit density
function of solutions to the Bethe equations on $C$. Let us denote
the endpoints of the contour $C$ by $\xi$ and $\bar{\xi}$,
$\xi=\lim_{N\to\infty} z_1$ and $\bar{\xi}=\lim_{N\to\infty}z_n$.

Taking the limit in the equation (\ref{b-eqn}) we obtain the
integral equation for $\rho(z)$:
\begin{equation}
\ln(z)=2H-\pi i\alpha +\frac{1}{2\pi}\int_C\theta(z,w)\rho(w)dw
+\int_{\xi}^z\rho(w)dw. \label{curve_eq}
\end{equation}
Here $z\in C$. Differentiating this equation with respect to $z$,
we obtain another, more convenient form of this equation:
\begin{equation}\label{eq-for-rhoz}
\rho(z)=\frac{1}{z}+\frac{1}{2\pi i}\int_C K(z,w)\rho(w)dw,
\end{equation}
where
\begin{equation}
K(z,w)=-i\partial_z\theta(z,w) =\frac{2\Delta-w}{1-2\Delta
z+zw}+\frac{w}{1-2\Delta w+zw}.
\end{equation}
The kernel $K$ is singular on two curves
\begin{eqnarray}
&& C_1=\{z=z_1(w)=\frac{1}{2\Delta-w}|\quad w\in C\}, \nonumber
\\
\\
&& C_2=\{z=z_2(w)=-\frac{1}{w}+2\Delta|\quad w\in C\}. \nonumber
\end{eqnarray}
and it can be written as
\begin{equation}
K(z,w)=-\frac{1}{z-z_1(w)}+\frac{1}{z-z_2(w)}.
\end{equation}

Notice that functions $z_2(w)=z_1^{-1}(w)$, i.e. $z_1$ and $z_2$
are inverse to each other.

If $\Delta=0$, coutours $C_1$ and $C_2$ coincide and the kernel is
zero.

\begin{conjecture}
If $\Delta\neq 0$, there are two cases:
\begin{itemize}
\item Conours $C$,$C_1$, $C_2$ do not intersect each other;

\item Contours $C$,$C_1$, $C_2$, {\it all} intersect at two points
$w_\pm=e^{\pm i\eta}$ which are the solutions of $w^2-2\Delta
w+1=0$ and are conjugate to each other (these two points may
coincide in the degenerate case). This is possible only if
$\alpha=1$.
\end{itemize}
\end{conjecture}

The integral equation on $\rho(z)$ can be rewritten as
\begin{equation}
\rho(z)=\frac{1}{z}+\frac{1}{2\pi i}\int_C
\frac{\rho(w)}{z-z_2(w)}dw -\frac{1}{2\pi i}\int_C
\frac{\rho(w)}{z-z_1(w)}dw. \label{int_eq}
\end{equation}
Notice that the contour $C$ can be deformed as long as it does not
intersect the curves $C_1$ and $C_2$.

The contour $C$ is defined by the condition that the form
$\rho(z)dz$ has purely imaginary values on the vectors tangent to
$C$
\begin{equation}\label{C-cond}
{\rm Re}(\rho(z)dz)\Bigr|_{z\in C}=0,
\end{equation}
and by the normalization condition on $\rho(z)$
\begin{equation}\label{rho-norm}
\frac{1}{2\pi i}\int_C\rho(z)dz=\alpha.
\end{equation}

The free energy per site is defined as
$$
f(H,V)=-\lim_{M,N\rightarrow\infty}\frac{\log Z_{N,M}}{NM}=
-\lim_{N\rightarrow\infty}max_{1\leq n\leq
N}\frac{\Lambda^{(n)}(N}{N}.
$$

The formula for eigenvaues $\Lambda^{(n)}(N)$ has two terms. Each
of them grow exponentially for large $N$. Generically, one of the
terms will dominate the other. Taking into account conjectures
about roots of Bethe equations in the thermodynamicallimit we
obtain the follwoing integral representation for the free energy
\cite{LW}:
\begin{eqnarray}
f&=&\min\Bigl\{\min_\alpha\Bigl\{E_1-H-(1-2\alpha)V-\frac{1}{2\pi
i} \int_C\ln(w-\frac{w^2-2\Delta w+1}{w-z})\rho(z)dz\Bigr\},
\nonumber
\\
&&\min_\alpha\Bigl\{E_2+H-(1-2\alpha)V-\frac{1}{2\pi i}
\int_C\ln(2\Delta-w+\frac{w^2-2\Delta
w+1}{w-z})\rho(z)dz\Bigr\}\Bigr\}, \nonumber
\end{eqnarray}
or
\begin{eqnarray}\label{free-en-form}
f(H,V)&=&\min\Bigl\{\min_\alpha\Bigl\{E_1-H-(1-2\alpha)V-\frac{1}{2\pi
i} \int_C\ln(\frac{b}{a}-\frac{c^2}{ab-a^2 z})\rho(z)dz\Bigr\},
\nonumber
\\
&& \min_\alpha\Bigl\{ E_2+H-(1-2\alpha)V-\frac{1}{2\pi i}
\int_C\ln(\frac{a^2-c^2}{ab}+\frac{c^2}{ab-a^2
z})\rho(z)dz\Bigr\}\Bigr\}, \nonumber
\end{eqnarray}
where $\rho(z)$ is determined bythe integral equation
(\ref{eq-for-rhoz}) with conditions (\ref{C-cond}) and
(\ref{rho-norm}).

\section{The Free Energy in the Free-Fermion Case}
\label{a_free_energy_free} The free-fermion curve in the space of
parameters of the 6-vertex model is $\Delta=0$. Since in this case
$K(z,w)=0$ the integral equation of $\rho$ immediately implies
$\rho(z)=1/z$. The contour $C$ in this case is an arc
\begin{equation}
z=e^{2H}e^{i \theta},
\end{equation}
where $\theta\in [-\pi\alpha,\pi\alpha]$.

Let us assume that $a>b$. The formula (\ref{free-en-form}) for the
free energy with fixed polarization $\alpha$ gives
\begin{equation}
f_\alpha=E_1-H-(1-2\alpha)V-\frac{1}{2\pi}
\int_{-\pi\alpha}^{\pi\alpha}\ln\Biggr|\frac{b}{a}
-\frac{a^2+b^2}{ab-a^2 e^{2H}e^{i\theta}}\Biggl|d\theta,
\end{equation}

Minimizing this expression in $\alpha$ we get the equation
\begin{equation}
\cos(\pi\alpha)=\frac{a^2(e^{2H+2V}-e^{-2H-2V})+b^2(e^{-2H+2V}-e^{2H-2V})}
{2ab(e^{2V}+e^{-2V})}
\end{equation}
This defines the critical value $0\geq \alpha\geq 1$ when the
absolute value of the r.h.s is not greater then $1$. Otherwise
$\alpha=0$ or $\alpha=1$.

\begin{proposition} The free energy can be
written as the following double integral:
\[
f=min_{0\leq \alpha\leq 1}
f_\alpha=-\frac{1}{4\pi^2}\int_0^{2\pi}d\theta\int_0^{2\pi}d\varphi
\ln\Bigl|a_1-a_2 e^{i(\theta+\varphi)}+b_1 e^{i\theta} +b_2
e^{i\varphi}\Bigr|.
\]
\end{proposition}
\begin{proof}The proof is computational. For the double integral in
question we have:

\begin{eqnarray}
&& \int_0^{2\pi}d\theta\int_0^{2\pi}d\varphi\ln\Bigl(a_1+b_1
e^{i\theta}+(b_2 -a_2 e^{i\theta})e^{i\varphi}\Bigr)= \nonumber
\\
&& 2\pi\int_{-\theta_*}^{\theta_*}d\theta\ln(a_1+b_1 e^{i\theta})
+2\pi\int_{-\pi+\theta_*}^{\pi-\theta_*} d\theta\ln(b_2+a_2
e^{i\theta}), \nonumber
\end{eqnarray}
where
\begin{eqnarray}
&& \theta_*=0\qquad {\rm if}\qquad |a_2-b_2|>a_1+b_1, \nonumber
\\
&& \theta_*=\pi\qquad {\rm if}\qquad |a_1-b_1|>a_2+b_2, \nonumber
\\
&& \cos(\theta_*)=\frac{a_2^2+b_2^2-a_1^2-b_1^2}
{2(a_1b_1+a_2b_2)}\,,\qquad {\rm otherwise}. \nonumber
\end{eqnarray}
The last equality holds if $|a_2-b_2|\leq a_1+b_1$ and
$|a_1-b_1|\leq a_2+b_2$.

We can rewrite $f$ as
\begin{equation}\label{s-int}
-8\pi^2 f=8\pi^2\ln\Bigl(\frac{a_1+b_1+|a_1-b_1|}
{2}\Bigr)+2\pi\int_{-\pi+\theta_*}^{\pi-\theta_*}
d\theta\ln\Biggl|\frac{b_2+a_2 e^{i\theta}} {a_1-b_1
e^{i\theta}}\Biggr|^2.
\end{equation}
We note that the argument of the logarithm under the integral is
greater than $1$ for any $\theta$ in the domain of integration.

Taking into account the parameterization of Boltzmann weights by
electric fields $H$ and $V$ it is easy to see that (\ref{s-int})
together with the equation for $\theta^*$ coincide with the
formula for the free energy derived from the Bethe Ansatz.

\end{proof}

The double integral formula  can be obtained directly from the
Pfaffian solution of the dimer model which maps to the 6-vertex
model at $\Delta=0$.

\end{document}